\documentstyle[12pt,epsfig,aaspp4,figwithcaption]{article}
\newcommand{\etal}{et~al.}
\newcommand{\Msun}{\ifmmode{M_\odot}\else$M_\odot$~\fi}
\newcommand{\kms}{\hbox{km\thinspace s$^{-1}$}}
\newcommand{\kmsMpc}{\hbox{km\thinspace s$^{-1}$\thinspace
Mpc$^{-1}$}} \newcommand{\TSPH}{{\sc TreeSPH}}
\newcommand{\fhot}{{f_{\rm hot}}} \newcommand{\disk}{{\rm disk}}
\newcommand{\infall}{{\rm inf}} \newcommand{\cold}{{\rm cold}}
\newcommand{\dm}{{\rm DM}} \newcommand{\jt}{{\tilde\jmath}}

\begin{document}
\slugcomment{to appear in Astrophysical Journal}
\title{Formation of Disk Galaxies:\\ Feedback and the Angular Momentum
problem} \author{Jesper Sommer-Larsen, Sergio Gelato and Henrik
Vedel\altaffilmark{1}} 
\affil{ Theoretical Astrophysics Center,
Juliane Maries Vej 30, DK-2100 Copenhagen {\O}, Denmark}
\altaffiltext{1}{Current Address: Dansk Meteorologisk Institut,
Lyngbyvej 100, DK-2100 Copenhagen {\O}, Denmark}

\begin{abstract}
When only cooling processes are included, smoothed-particle 
hydrodynamical simulations
of galaxy formation in a Cold Dark Matter hierarchical clustering
scenario consistently produce collapsed objects that are deficient in
angular momentum by a factor of about~25 relative to the disks of
observed spiral galaxies.  It is widely hoped that proper allowance
for star formation feedback effects will resolve this discrepancy.  We
explore and compare the effects of including two different types of
feedback event: uniform reheating of the entire Universe to $5\times
10^5$~K at a redshift~$z\sim 6$, and gas blow-out from pregalactic gas
clouds comparable to present-day dwarf galaxies at a somewhat lower
redshift.  We find that blow-out is far more successful than early
reheating, and that it may even be sufficient to solve the angular
momentum problem. We see indications that the remaining angular
momentum deficit (a factor of~5 in the blow-out models) is due to
current limitations in the numerical method (SPH) used in our and other
authors' work.  Our most successful models distinguish themselves by
the fact that a large fraction of the gas is accreted very gradually,
in a cooling flow from a surrounding hot phase, rather than by mergers
of massive cold clumps.
\end{abstract}
 
\section{Introduction}

The formation of galactic disks is one of the most important unsolved
problems in astrophysics today. In the currently favored hierarchical
clustering framework, disks form in the potential wells of dark matter
halos as the baryonic material cools and collapses dissipatively.  It
has been shown (Fall \& Efstathiou \cite{FE80}, hereafter FE80) that
disks formed in this way can be expected to possess the observed
amount of angular momentum (and therefore the observed spatial extent
for a given mass and profile shape), but only on condition that the
infalling gas retain most of its original angular momentum.

Numerical simulations of this collapse scenario in a cosmological
context 
(e.g., 
Navarro \& Benz \cite{NB91},
Navarro \& White \cite{NW94},
Navarro, Frenk, \& White \cite{NFW95}), 
however,
have so far consistently indicated that when only cooling processes are
included the infalling gas loses too much angular momentum (by over
an order of magnitude) and the resulting disks are accordingly much
smaller than required by the observations.
This discrepancy is known as the {\em angular momentum problem} of
galaxy formation.
It arises from the fact that gas cooling
is very efficient at early times, leading to the rapid condensation of
gas clouds which are then promptly slowed down by dynamical
friction against the surrounding dark matter halo before they
eventually merge to form the central disk.
A mechanism is therefore needed that prevents, or at least delays,
the collapse of protogalactic gas clouds and allows the gas to
preserve a larger fraction of its angular momentum as it settles into
the disk.
(Such a mechanism is also helpful in solving the {\em overcooling
problem}, namely the observation (White \& Rees \cite{WR78})
that cooling is expected to be so efficient at early times that most
of the gas should have been converted to stars well before the
assembly of present-day galactic disks.)
Weil, Eke, \& Efstathiou (\cite{WEE98}, hereafter WEE98) 
have shown that if the
cooling is suppressed (by whatever means), numerical simulations can
indeed yield more realistically sized disks.
The physical mechanism by which cooling is suppressed or
counteracted, however, is still uncertain.

An early candidate was the extragalactic ionizing ultraviolet and
soft X-ray (UVX) background which is thought to permeate the Universe at
least at redshifts $z\la 3$--$4$ (with quasars as an important source)
and maybe considerably earlier.
Such a field provides a natural explanation of the
observed low optical depth in neutral hydrogen (Gunn \& Peterson
\cite{GP65}, Efstathiou \cite{Ef92}) and of the
properties of the Lyman~$\alpha$ forest (Hernquist {\etal}
\cite{H.96}).
The effect of such a field on galaxy formation was explored by Vedel, 
Hellsten, \& Sommer-Larsen (\cite{VHS94}, hereafter VHSL94).
These authors concluded that the UVX background does indeed affect
the formation of both small and large galaxies.
However, the UVX background is not in itself sufficient to solve the angular
momentum problem: it is better at inhibiting the accretion of
low-density gas at low redshift than the formation of high-density
clouds at high redshift (when inverse Compton cooling is particularly
efficient and capable of counteracting the UVX heating, even if one assumes
that the UVX field is already strong at that time).
This has been pointed out among others by Navarro \& Steinmetz
(\cite{NS97}, hereafter NS97), 
who even find that the UVX field exacerbates
the angular momentum deficit. (This last detail is at variance with
the result of VHSL94. The broader fact that the UVX field alone does not
solve the angular momentum problem, however, is much less controversial.)

A different heating mechanism is therefore required.
In this paper we explore and compare two candidate mechanisms: uniform
reheating of the entire Universe at $z\sim6$ (as could have been
caused by energy feedback from a pregalactic phase of star formation),
and more conventional, localized
feedback in the form of gas blow-out by supernova explosions
in the first generation of collapsed subgalactic objects at a somewhat
later epoch. We also compare the results to those of purely
passive cooling simulations, with no heat sources other than the UVX
background.

We begin (in section~\ref{s:scenarios}) with some plausibility arguments and
estimates of relevant physical parameters for our two reheating
scenarios. 
In particular, we derive (\S~\ref{s:trhmax}) an upper bound on the
reheating temperature in the first scenario.
Section~\ref{s:methods} briefly presents the numerical
methods used, namely the numerical code and the model
parameters and initial conditions. The simulations themselves are
presented in section~\ref{s:simulations}, and the results analyzed in
section~\ref{s:results}. We summarize our conclusions in
section~\ref{s:conclusions}.

\section{Reheating scenarios}
\label{s:scenarios}

\subsection{Early reheating of the Universe}
\label{s:trhmax}

Consider the hypothesis that the Universe was reheated and reionized
at a relatively high redshift $z_{RH}$, when it was still fairly
homogeneous on galactic scales.
One way in which this could have come about is if at those redshifts
a fraction of the gas was had collapsed into a dense, cold phase
(cf. Tegmark {\etal} \cite{Te.97}) 
out of which a small amount of stars
formed. The more massive of these would quickly explode as supernovae,
releasing enough energy to reheat the medium more or less uniformly
to temperatures $T_{RH} \sim 10^5$~K, as we proceed to show.

Assuming that a fraction~$f_\star$ of the gas is turned into stars,
that each unit of mass in stars gives rise (over a short time scale of
a few million years) to $\nu_{SN}$ type~II supernovae
each of which releases an energy $E_{SN}\sim 10^{51}$~erg, the
energy released by supernovae per unit mass can be written as
\begin{equation}
f_{\star}\nu_{SN} E_{SN}.
\end{equation}
To account for radiative losses we further assume
that the released energy is turned into thermal energy
of the remaining ISM with some efficiency $\beta \le 1$.
One can easily show that the temperature of the remaining gas 
then increases to
\begin{equation}
\label{q:t-rh}
T_{RH} \simeq \frac{2}{3} \frac{\mu m_p}{k_B} \nu_{SN}
E_{SN} \frac{\beta f_{\star}}{1 - (1-R) f_{\star}} \simeq
2.4 \times 10^5 \beta \nu_{100} E_{51} 
\left(\frac{f_{\star}}{0.01}\right) \mbox{K},
\end{equation}
where $\mu$ is the molecular weight (which we take to be 0.6), 
$m_p$ is the mass of the proton,
$k_B$ is Boltzmann's constant,
$R$ is the stellar return fraction ($0 \le R \la 0.4$ for ``standard''
initial mass functions),
$\nu_{100} = (100 \Msun) \nu_{SN}$,  
and $E_{SN} = 10^{51} E_{51}$~erg.
In the last part of equation~(\ref{q:t-rh}) we have assumed 
$(1-R) f_{\star} \ll 1$.

We now give two reasons why $f_{\star}$ must be small, resulting in an
upper limit on $T_{RH}$.
First, most of the stars that form at $z_{RH}$ and are responsible for
reheating the gas out of which disk galaxies later condense are expected to
end up in the halos of these same galaxies (cf. Sommer-Larsen {\etal}
\cite{S.97}); as the mass of the
stellar halo of the Milky Way is only about
1\% of the total baryonic mass, $f_{\star} \la 0.01/(1-R)$.
For a Scalo (\cite{Sc86}) ``best fit'' initial mass
function (IMF), $\nu_{100} \sim 0.4$. 
So with $E_{51} \simeq 1$ one would expect, from
equation~(\ref{q:t-rh}), that $T_{RH} \la 10^5$~K. 
There are some indications (Kennicutt, Tamblyn \& Congdon
\cite{KTC94}, Sommer-Larsen \cite{So96}, Tsujimoto
{\etal} \cite{Ts.97}) 
that the global IMF is more top-heavy than Scalo's, implying
$\nu_{100} \sim 1$.
Using this, and setting $\beta = 1$, a firm upper limit to $T_{RH}$ is
$T_{RH} \la 5 \times 10^5$~K.
Note, however, that if the early IMF was extremely top-heavy
(and consequently $(1-R) \ll 1$)
the Milky Way stellar halo mass argument given above could be bypassed and
one could in principle have $f_{\star} \gg 0.01$ and $T_{RH} > 5
\times 10^5$~K.

The second, stronger argument is based on chemical evolution
considerations, in particular the fact that oxygen is ejected only by
type II supernovae ($M \ga 8\Msun$).
Beers \& Sommer-Larsen (\cite{BS95}) found that the disk of
the Milky Way extends down to very low metallicities: $[Fe/H] \la -2$,
or equivalently $[O/H] \la -1.5$ as $[O/Fe] \simeq 0.5$ in Galactic 
metal-poor stars. 
Let $M_O$ be the mass of oxygen 
produced and ejected by a type II supernova. 
Its value is fairly insensitive to the IMF: $M_O \simeq 2.0\Msun$ 
for a Scalo IMF and  $\simeq 2.6\Msun$
for a Salpeter IMF (cf. Pagel \cite{Pa97}). 
These numbers are consistent with the observationally
inferred value for SN 1987A.
The oxygen abundance is then given by:
\begin{equation}
\label{q:zi-o}
Z_i(O) = \frac{f_{\star}}{1 - (1-R) f_{\star}} \nu_{SN}  M_O
\simeq f_{\star} \nu_{SN} M_O,
\end{equation}
where again we have made use of $(1-R) f_{\star} \ll 1$ in the last
 part of
equation~(\ref{q:zi-o}).
It then follows that
\begin{equation}
f_{\star} \simeq \frac{Z_i(O)}{\nu_{SN} M_O} = 0.016 \left(\frac{Z_i(O)}{10^{-1.5}
Z_{\odot}(O)}\right) \nu_{100}^{-1} \left(\frac{M_O}{2 \Msun}\right)^{-1}.
\end{equation}
Inserting this into equation~(\ref{q:t-rh}) yields
\begin{equation}
T_{RH} \simeq \frac{2}{3} \frac{\beta \mu m_p E_{SN} Z_i(O) }{k_B M_O} 
=
3.8 \times 10^5 \beta E_{51} \left(\frac{Z_i(O)}{10^{-1.5} Z_{\odot}(O)}\right)
\left(\frac{M_O}{2 \Msun}\right)^{-1} ~\mbox{K},
\end{equation}
for $\mu$ = 0.6 and $Z_{\odot}(O)$ = 0.01 (e.g., Pagel \cite{Pa97}). 
This expression does not
depend on the values of $\nu_{SN}$ and $R$, so it is a considerably
more robust estimate of $T_{RH}$ than what is obtained with the
Galactic stellar halo argument above. 
The existence of a very metal-weak tail of the Galactic disk has
been questioned (Ryan \& Lambert \cite{RL95}, Chiba \& Yoshii
\cite{CY97}), but there are
other indications that the typical abundance produced by a
(hypothetical) early phase of star formation (``population~III'') is
very low.
In Lyman~$\alpha$ clouds with neutral 
hydrogen column density $N(H I)$ as low as $10^{14.5}\,\mbox{cm}^{-2}$,
carbon is detected but at a low level: $[C/H] \sim -2.5$ (Songaila \&
Cowie \cite{SC96}, Lu, Sargent, \& Barlow \cite{LSB97}). 
With $[C/O] \simeq -0.5$, as detected in metal-weak
stars as well as in metal-weak extragalactic H II regions (e.g., Pagel
\cite{Pa97}), this corresponds to $[O/H] \sim -2$, similar to
what is obtained from the metal-weak 
thick disk tail argument above. 
There are even indications (Ostriker \& Gnedin \cite{OG96},
Gnedin \& Ostriker \cite{GO97}, Lu {\etal} \cite{L.98})
that population~III enrichment is at least 1.5~dex lower than this.

At very high redshifts inverse Compton cooling dominates. The cooling
rate is given by (Ikeuchi \& Ostriker \cite{IO86})
\begin{equation}
\label{q:compton}
\Lambda_C = 5.41 \times 10^{-36} n_e T (1+z)^4 \mbox{erg s$^{-1}$ cm$^{-3}$},
\end{equation}
where $n_e$ is the number density of free electrons in units of
cm$^{-3}$. 
For a fully ionized, primordial plasma the inverse Compton cooling
timescale is given by 
\begin{equation}
t_C = \frac{\cal{E}}{\dot{\cal{E}}_C} = \frac{3 n_e k_B T}{(X+1) \mu \Lambda_C} 
\simeq 10^9 \left(\frac{7}{1+z}\right)^4 \mbox{yr}
\end{equation}
which is independent of density and temperature.
($X$ is the Hydrogen mass fraction; we assume $X=0.76$ in our
numerical estimates.)
For an $\Omega=1$, $H_0=50$~{\kms} cosmology $t_C$ is comparable to the
Hubble time at $z=z_{eq}\simeq 7$.
For $z_{RH} \ga z_{eq}$ reheating is therefore ineffective.
We consequently choose $z_{RH} < z_{eq}$.

\subsection{Gas blow-out from pregalactic clouds}

An alternative to this diffuse reheating scenario is to assume that
larger condensations do form early on, but that star formation
activity is strong enough to expel the gas from these regions 
before they merge into larger objects (Dekel \& Silk \cite{DS86}).
The gas then needs to cool and condense again,
now more slowly and gradually since the mean density
is lower. The most significant change is that the disk grows in a
gentle cooling flow from a dilute, hot ($T \sim 10^6$~K) corona
rather than through mergers of inspiralling high-density
clumps.
The primary feedback process is supernova
explosions, but stellar winds and UV radiation from massive stars also
contribute to reheating the interstellar medium (Hellsten \cite{He95}).

In this scenario, reheating occurs in a more patchy manner and
radiative cooling is the main competing effect.
Unfortunately the balance between heating and cooling is very
difficult to establish from first principles, since it depends
strongly on the manner in which the energy from the stars is
deposited into the ISM. Supernovae are thought to be largely
responsible for maintaining the multiphase structure of the ISM.
This is an area of ongoing research, and best discussed in a separate
paper. Sommer-Larsen, Hellsten \& Vedel (\cite{SHV99}) use a simplified
treatment of the ISM as a two-phase medium composed of a warm ($T \sim
10^4$~K) and a hot ($T \sim 10^6$~K) phase to show that
supernova-driven hot gas outflows can be very efficient during the
early stages of galaxy formation when the metallicity of the gas is low.

To illustrate why the multiphase structure of the ISM cannot be
ignored for a proper treatment, let us consider a schematic model
similar to that used by various authors (e.g., Katz
\cite{Ka92}, Navarro \& White \cite{NW93}, Mihos \&
Hernquist \cite{MH94}) to represent star formation and
feedback in numerical simulations of galaxy formation.
Let the star-formation rate in a dense gas cloud be given by
\begin{equation}
\frac{dM_{\star}}{dt} = \frac{M_{gas}}{t_{\star}},
\end{equation}
where 
$t_{\star} = \tau t_{ff}$, $t_{ff} = (4 \pi G \rho_{gas})^{-1/2}$,
and $\tau \ge 1$ is an inverse star-formation efficiency parameter.
The rate of SN explosion energy input into the inter-stellar medium (ISM) is
then given by
\begin{equation}
\dot{E}_{heat} = \frac{dM_{\star}}{dt} \nu_{SN} E_{SN},
\end{equation}
in which $\nu_{SN}$ and $E_{SN}$ are defined as in equation~(\ref{q:t-rh}).
The rate of radiative cooling per unit volume is written in terms of
the {\em cooling function} $\Lambda_{cool}(T)$ as
\begin{equation}
\dot{\cal{E}}_{cool} = \Lambda_{cool}(T) n_H^2.
\end{equation}
It is straightforward to show
that the ratio of the rate of SN energy input to the rate of cooling
in the gas cloud is 
\begin{equation}
q \equiv \frac{\dot{E}_{heat}}{\dot{E}_{cool}} = 
\frac{\nu_{SN} E_{SN} (4 \pi G)^{1/2} m_p^{3/2}}{\tau 
\Lambda_{cool}(T) n_H^{1/2} X^{3/2}} 
= 1.5 \frac{\nu_{100} E_{51}}{\tau \Lambda_{-23} n_H^{1/2}} ,
\end{equation}
where $\nu_{100} = (100\Msun) \nu_{SN}$,  
$E_{SN} = 10^{51} E_{51}\,\mbox{erg}$, 
$\Lambda = 10^{-23}\Lambda_{-23}\,\mbox{erg cm$^{-3}$ s$^{-1}$})$,
and $n_H$ is the hydrogen number density in units of cm$^{-3}$.

The balance $q$ between the rates of heating and cooling depends
fairly sensitively 
on the mean effective density $n_H$ of the gas. Katz
(\cite{Ka92}) found in his simulations that if the gas is
treated as a single-phase medium (of relatively high mean density
in the star-forming regions) that is heated more or less uniformly
by feedback processes, then $q\ll1$ and the heat is quickly radiated
away, resulting in no appreciable blow-out.
On the other hand, if the supernova energy is unevenly distributed
and predominantly serves to heat a low-density gas phase, then it can only
be radiated away over a much longer time scale, and the effect of
multiple supernova explosions can be additive.

Other mechanisms for local reheating have been proposed.
For example, the radiation from a quasar located in the host
protogalaxy could heat the surrounding gas.
However, one can show that photoionization cannot raise the gas
temperature above $T_{QH} \la 10^5$~K for any reasonable choice of
quasar spectrum (Hellsten, private communication).
Compton heating has been discussed by Krolik, McKee, \& Tarter
(\cite{KMT81}). It might be important relatively close ($d \la
10$~kpc) to the quasar. More important, however, might be the effect
of quasar winds, as discussed recently by Silk \& Rees
(\cite{SR98}). 
Note, however, that the comoving density of quasars peaks quite late,
at $z\sim 2$; it remains to be seen whether this is early enough for
the reheating to be effective.

\section{The code and the initial conditions}
\label{s:methods}

\subsection{The code}
\label{s:code}

We use the gridless Lagrangian $N$-body and Smoothed Particle
Hydrodynamics code {\TSPH} described in VHSL94, with only a
minor change in the method by which the energy equation for the gas is
solved. Our {\TSPH} code is modeled after that of Hernquist \& Katz
(\cite{HK89}).

We include gas cooling and heating terms as in VHSL94. The heating
corresponds to a redshift-dependent, homogeneous and isotropic UVX
background field. We assume a rather hard (spectral index $-1$) field
\begin{equation}
J_{\nu}(z) = J_{-21}(z) \times 10^{-21}
\left(\nu\over\nu_L\right)^{-1}
\,\hbox{erg}\,\hbox{cm}^{-2}\,\hbox{sr}^{-1}\,\hbox{Hz}^{-1}\,\hbox{s}^{-1},
\end{equation}
where $\nu_L$ is the Lyman limit frequency,
with the redshift-dependent normalization
\begin{equation}
J_{-21}(z) = { 10 \over 1 + \left[5/(1+z)\right]^4 }
\end{equation}
of Efstathiou (\cite{Ef92}).
For greater realism one could build on the detailed study of Haardt \&
Madau (\cite{HM96}), but our adopted background field is
quantitatively not too dissimilar from theirs (once allowance is made
for our simplified spectral shape), at least at redshifts~$z\la 3$,
and should be adequate for the level of detail we can represent in
the simulations.
The code also incorporates inverse Compton cooling
(equation~[\ref{q:compton}]), which is also
explicitly redshift-dependent.

Due to the impossibility of properly representing the multiphase
structure of the ISM in an SPH simulation  at the level
of resolution we can afford to run and to the lack of a clear
physical basis (or even of a reliable empirical calibration) for
schemes such as the kinetic energy feedback term of
Navarro \& White (\cite{NW93}), we do not attempt to follow
star formation and feedback in the manner pioneered by Katz
(\cite{Ka92}). Instead, we mimic feedback events occurring at
well-defined epochs by instantaneously altering the state of the
gas at that time, then continuing the simulation. This is adequate
for our present purpose, which is to gain a qualitative insight into
the effect of varying the epoch and manner of reheating rather than to
pursue the much more ambitious and long-term goal of a fully realistic
simulation. 

The simulations use artificial viscosity parameters $\alpha_v=0.9$,
$\beta_v=2\alpha_v$. The smoothing length of each SPH particle is
adjusted so as to keep the number of neighbors close to~50.

\subsection{The initial conditions}
\label{s:ic}

Our initial conditions are based on a standard ($\Omega=1$,
$\Lambda=0$) Cold Dark Matter (CDM) model with Hubble constant
$H_0 = 100h\,\kmsMpc = 50\,\kmsMpc$. On the scales of interest to
us the effective index of the power spectrum is approximately~$-2$:
$P(k) \propto k^{-2}$. Following Eke, Cole, \& Frenk
(\cite{ECF96})
we normalize the spectrum to $\sigma_8(z=0)=0.5$,
where as customary $\sigma_8^2$ is the mass variance within spheres of
comoving radius $8h^{-1}$~Mpc, extrapolated from the linear regime of
perturbation growth.

We begin by performing a large-scale simulation within a sphere of
comoving radius 40~Mpc. Individual halos are then selected from the
final state and individually sampled at higher resolution.
The original large-scale simulation is used to provide a tidal
field acting on the resampled halos, which are evolved individually in
a second round of simulations.

Approximately $2.5\times 10^5$ particles are initially placed on a
cubic lattice 
within that sphere. Position and velocity perturbations
are then applied according to the Zel'dovich (\cite{Ze70})
approximation.
The perturbations consist of the superposition of $N_k \sim
4\times10^4$ plane waves sampling a Gaussian random field with
variance given by the power spectrum.
Following Navarro \& White (\cite{NW94}) we use an equal number of
waves per logarithmic interval in $k$-space. The phases of these waves are
random and only wavenumbers between the fundamental and Nyquist wavenumber
of the lattice are included.
The initial redshift (which determines the amplitude of the initial
perturbations) is $z_i\simeq 18$.

The evolution of this system to $z=0$ is then computed using a tree code. Only
gravitational forces are included in this first simulation.
In the final state we identify four virialized, isolated halos with
circular velocities between 200 and 260~{\kms} in the innermost 20~Mpc of the
simulation.
We expect that a significant fraction of such halos should host disk
galaxies similar to the Milky Way since the circular velocities are
in the same range and our halos were chosen to lie ``in the field'',
away from larger concentrations of mass.
We adopt the customary working definition of the virial radius as the
radius~$r_{200}$ of a sphere enclosing a mean density of 200 times the
critical cosmic value.
Tracing the particles in these halos back to the initial conditions,
we find that they all come from regions that fit within spheres of
comoving radius $\sim 3$~Mpc.

Each of these spheres at $z\simeq18$ is then resampled with a lattice
4 times finer in each spatial dimension. Each sphere contains about
7000 points of this new lattice. We assign one dark matter (DM)
particle and two SPH particles to each of these points. The SPH
particles are positioned at random within one gas gravitational softening
length of their parent DM particle.
Note that at this redshift the DM particles are spaced by about
14~kpc, so that each SPH particle is rather closely associated with
its parent DM particle.
Table~\ref{t:np} lists the precise number of SPH and DM particles in
each sphere.
We adopt a baryonic mass fraction $\Omega_b=0.05$, consistent with
nucleosynthesis constraints ($0.01 h^{-2} \la \Omega_b \la 0.02
h^{-2}$). This leads to masses of $1.1\times 10^9\,\Msun$ for the
DM and $2.9\times 10^7\,\Msun$ for the SPH particles.
The SPH particles are assigned an initial thermal energy corresponding
to a temperature $T_i\simeq 100$~K.
In order to include small-scale power that could not be sampled in the
first simulation, we add shorter-wavelength plane waves in a way that
preserves an equal number of waves per interval in~$\log k$.

We use stored intermediate results from the large cosmological
simulation to provide a time-dependent tidal field acting on the
resampled spheres.
This is achieved by treating the original particles outside the
resampled sphere as passive, interpolating their positions between
successive snapshots of the original simulation and incorporating them
into the particle tree that the code constructs on each step for the
evaluation of gravitational forces.
Gravitational interactions between particles are softened
according to the prescription of Hernquist \& Katz (\cite{HK89}), with
softening lengths 3 kpc for the gas particles, 10 kpc for the dark matter 
particles, and 40~kpc for the ``passive'' dark matter particles that
provide the tidal field from the original large-scale simulation. The
gravitational softening lengths are kept 
constant in physical units throughout the evolution of the system.

\section{The simulations}
\label{s:simulations}

We carry out three sets of four simulations each. In what follows,
the symbols $S1$, $S2$, $S3$, and $S4$ correspond to the four high-resolution
spheres, and the subscripts $PA$, $RH$, and $BO$ to the three sets of
simulations. These subscripts stand respectively for ``passive'',
``reheating'', and ``blow-out''.

In the $PA$ simulations, no stellar feedback is applied to the
thermal history of the gas. The only non-adiabatic effects present
(aside from the usual shock-capturing artificial viscosity) are
radiative and inverse Compton cooling, and heating through
photoionization by the background UVX field.
These simulations are meant to extend the results of VHSL94 to more
realistic initial conditions than the uniform top-hat these authors
used, as well as to provide a basis for assessing the effectiveness of
our reheating experiments.

In the $RH$ simulations, the gas is reheated wholesale to a
temperature of $5\times 10^5$~K at a redshift $z=6$.
(We have also performed a test simulation in which this reheating occurs at
$z=18$ rather than $z=6$, and found as expected that the heat was
dissipated very quickly by inverse Compton cooling. The final result
was essentially indistinguishable from that of the corresponding
``passive'' simulation.)
Apart from this instantaneous reheating event, the system is evolved
passively, exactly as in the $PA$ runs.

In the $BO$ simulations, all cold ($T < 3\times 10^4$~K), 
relatively dense ($n_H > 10^{-3.5}
\mbox{cm$^{-3}$}$) gas clumps are identified in the
simulation at $z=2.4$ and their particles instantaneously
redistributed uniformly within spherical shells between radii
$r_{\rm in}=\alpha r_0$ and $r_{\rm out}=2\alpha r_0$ around the
center of mass of each clump. In other words, the gas is blown out but
not blown away from the systems.
Here $r_0 \sim 50~\mbox{kpc}$ is the virial
radius of the largest progenitor halo to the final object at that
redshift ($z = 2.4$), and $\alpha$ a tuning factor of order unity.
The results we present here were obtained with $\alpha=1.2$, but
tests we have performed suggest that the results are not very sensitive
to the precise value of this parameter. 
For robustness $r_0$ is not measured directly from the state of the
simulation at $z=2.4$ but estimated from $V_{200}$ for the final halos
(at $z=0$) using the isothermal sphere approximation (cf. Mo, Mao, \&
White \cite{MMW98})
\begin{equation}
\label{q:r-zero}
r_0 = \frac{V_{200}}{10 H(z)} = 400 \left(\frac{V_{200}}{200\,\mbox{km s}^{-1}}\right) 
(1+z)^{-3/2}\, \mbox{kpc},
\end{equation}
where $H(z)$ is the Hubble parameter at redshift $z$ and the last part
of equation~(\ref{q:r-zero}) follows from our choice of cosmology.
The temperature of the ``blown out'' gas was not changed and, for a 
given gas cloud, all the ``blown out'' gas was given the same
velocity, the center of mass velocity of the parent gas
cloud. This was done to ensure that no extra angular momentum or
thermal energy was added to the simulations (the angular momenta
of the original cold gas clouds were insignificant).
As in the $RH$ case, no form of feedback is applied apart from this
instantaneous event.
The redshift $z=2.4$ is appropriate in our case because our
simulations at that time contain a number of gas clumps of the
requisite size (namely the largest size of object for which blow-out
can reasonably be expected to be significant), 
and these are about to merge into larger structures.
The picture in our simulations presents a striking resemblance to the
Hubble Space Telescope observation by Pascarelle {\etal}
(\cite{P.96}) of a group of small galaxies at that same
redshift, apparently in the process of merging into a larger object.
In a more realistic implementation (which may well be required if
one were to do such simulations at higher resolution) the blow-out
would occur continuously rather than at a single
point in time as in the schematic model we study here.
Note that our choice of redshift is tied to the formation history of
the galaxies we are studying, and might therefore be sensitive to the
choice of cosmological model and to the mass of the final halo.

\section{Results}
\label{s:results}

\subsection{Disk masses, sizes, angular momenta}

Table~\ref{t:mrv} (which is closely modeled after Table~1 of NS97)
presents some global properties of the final collapsed objects
in our twelve simulations. The first column is the simulation label,
as described in the previous section. Columns 2, 3, and~4 respectively
hold the virial mass $M_{200}$, the virial radius $r_{200}$, and the
circular velocity $V_{200}$ at the virial radius.
The numbers $N_{\rm gas}$ and~$N_{\rm DM}$ of gas and DM particles
inside the virial radius are in columns 5 and~6, and the corresponding
masses $M_{\rm gas}$ and~$M_{\rm DM}$ in columns 7 and~8.
Columns 9, 10, and~11 show the number of gas particles $N_\disk$,
the corresponding baryonic mass~$M_\disk$, and the ratio $M_\disk
/\Omega_bM_{200}$ for the cold disk present at the end of each
simulation.
We define this disk as consisting of all cold ($T < 3\times 10^4$~K) gas
particles within $r_{\rm max}=10$~kpc from the (iteratively
determined) center of mass of the disk.
The results remain virtually unchanged if we adopt $r_{\rm max}=5$~kpc
instead.
For the $BO$ models we used $r_{\rm max}$ values of 20 and
10~kpc, respectively, since it turns out that larger disks can form
in these models.

In all the simulations, only about 40--60\% of the gas mass inside
the virial radius has cooled and settled into the disk; this low
fraction is largely due to the presence of the UVX field (VHSL94,
NS97). We find no systematic trend in cooled mass fraction between
the three sets of simulations.
For a given set of initial conditions, the final virial mass $M_{200}$
is very nearly the same for all three runs.
The gas mass $M_{\rm gas}$ fluctuates a little more, but without
exhibiting any clear systematic trend.
The scatter in the cooled central disk gas mass is larger still, but
(perhaps remarkably) 
does not exhibit any clear systematic trend either; a much larger number of
realizations would probably be needed if we wanted to find a dependence.

How large are the cold disks, and how much angular momentum do they
contain?
We need to answer this question both for the simulation results and
for real galaxies. For the latter we start from existing catalogues by
Mathewson, Ford, \& Buchhorn (\cite{MFB92}) and by Byun
(\cite{By92}). The full catalogues are diameter-limited, but
one of us (JSL) has carefully extracted a subsample of 203 late-type
(Sb--Sd) galaxies in such a way as to compensate for the selection
bias of the full dataset.
We use as a measure of the linear sizes of the observed disks their
inclination-corrected $I$-band exponential disk scale lengths $b_I^0$.
For consistency with the numerical simulations we assume $h=0.5$ in
converting from redshifts to linear distances.
It turns out that for our sample $b_I^0 \propto V_c^\eta$ (with some
scatter), where $V_c$ is the amplitude of the flat part of the
rotation curve and $\eta \simeq 1$.
We cannot directly measure the specific angular momenta of these
disks, but we can estimate them from the relation
\begin{equation}
j = 1.68 b V_c,
\end{equation}
which holds exactly for a purely exponential disk truncated at
$R_t=4b$
with a perfectly flat rotation curve and no bulge
contribution.
The radial cutoff corresponds to the observation by van der Kruit \&
Searle (\cite{KS82}) that the disks of spiral galaxies have
an edge at that radius.
It follows from $b \propto V_c^\eta$ that $j \propto V_c^{1+\eta}$.
In much of the rest of this paper, we express results in terms of a
normalized specific angular momentum $\jt \equiv j/V_c^2$.

For the simulations, it is easier to determine $j$ directly (since we
have full knowledge of the particle distribution in phase space),
but in many cases harder to assign an exponential scale length~$b$ due
to resolution limitations.
We tentatively estimate $b$ as
$\langle R^2\rangle^{1/2}/1.94$,
which again holds exactly for a perfect exponential disk truncated at
$R_t=4b$. (Here $\langle R^2\rangle$ is the mass-weighted average of the
square of the cylindrical radius~$R$.)
We measure $V_c$ at $R=30$~kpc, where the rotation curves turn out to
be approximately flat. The values of $V_c$ and $b$ are tabulated
respectively in columns 2 and~3 of table~\ref{t:j}.

In figure~\ref{f:b-vc} we plot the exponential disk scale length~$b$ as a function of
circular speed~$V_c$, both for our simulation results (filled
triangles represent the $PA$ runs, squares the $RH$ runs, circles the
$BO$ runs) and for the observational data.
One can clearly see from this figure that the simulated disks are
far too centrally concentrated for their circular speed. 
The situation is particularly
bad for the passive simulations (whose scale lengths~$b$ are too small
by an order of magnitude); by contrast, one of the blow-out runs
is almost acceptable. The early reheating runs are intermediate.

Figure~\ref{f:j-b} shows the relation between $j$ and~$b$ for the simulations.
(The values of $j$ are listed in column~5 of table~3.)
It is closer to quadratic ($j \propto b^2$) than
linear. Sommer-Larsen, Vedel, \& Hellsten (\cite{SVH98})
interpret this as a consequence of the gravitational softening length
being comparable to or larger than the extent of the disk.
This indicates that the problem is actually even worse than figure~1
would imply, since the values of $b$ from the simulations are 
artificially inflated as an effect of gravitational softening.

In figure~\ref{f:jdisk} we plot the values of $j/V_c^2$ versus $V_c$ for the
simulations. The solid horizontal line corresponds to the mean
from the observational data set, with dotted and dashed lines
corresponding to one and two standard deviations from this mean.
For the passive simulations, the angular momentum deficit is highly
significant, and amounts to a factor of about 25.
This amount is broadly consistent with the findings of NS97. 
(Navarro, Frenk, \& White (\cite{NFW95}) found a smaller factor
of about 5, but as discussed in NS97 this is probably due to
resolution problems.
The simulations we present here have nearly the same resolution
as those of NS97.)
Our ``reheating'' and ``blow-out'' simulations clearly reduce the gap
with the observations; the blow-out models are particularly
encouraging. The deficit for the latter
is only about a factor~5, which can perhaps
be accounted for by limitations in the numerical method.
A detailed examination of the results of WEE98 (see especially their
figures $1c$ and~14) reveals that their stellar disks also lie
below the observations by a similar factor. We are inclined to agree
with them in considering this a success.

Besides the values of $V_c$, $b_\disk$, and $j_\disk$
already mentioned, table~\ref{t:j} lists our computed values for the
dimensionless spin parameter $\lambda \equiv J|E|^{1/2}/GM^{5/2}$
(column~4) evaluated at the {\em infall radius} $r_\infall$
(column~6). This radius is defined by $M_{\rm DM}(r_\infall)/(1-\Omega_b) =
M_\disk/\Omega_b$, and is of order 100~kpc in all our runs.
It represents the characteristic radius, at the present time, of the
dark matter originally associated with the amount of gas currently in
the disk.
Evaluating $\lambda$ at $r_{200}$ does not significantly affect the
results.
The spin parameters for our halos are consistent with the theoretical
prediction of a broad distribution around $\lambda \sim 0.05$ 
(Barnes \& Efstathiou \cite{BE87},
Heavens \& Peacock \cite{HP88}).
The angular momentum deficit in our disks is therefore unlikely to be
due to our having chosen an unrepresentative set of parent halos.

The azimuthally averaged disk surface density profiles
are shown in figure~\ref{f:sd-r-bo} for the $BO$ runs.
The disks are close to exponential over a reasonable range of radii
(out to at least 4~kpc), with scale lengths of order 1--2~kpc.
Figure~\ref{f:vc-r-bo} shows the rotation curves, recomputed from the particle
distribution at $z=0$ under the assumption of negligible gravitational
softening, for the same four runs. Consistent with our previous
observation that the angular momentum deficit has been reduced but not
completely eliminated, there is still a hint of an
excessive central concentration of mass in some of the runs, but the
general behavior appears reasonably realistic otherwise.

The most extended disk found in our simulations is that of $S4_{BO}$.
We show an edge-on view in figure~\ref{f:z-y-s4bo}, and a face-on view (with arrows
to represent the velocity field) in figure~\ref{f:y-x-s4bo}.
This disk extends beyond 10~kpc and exhibits a distinct warp in the
outer regions. A spiral arm may even be discerned in the face-on view.

\subsection{Analysis of the infall process}

We distinguish between a ``cold'' ($T < 3\times 10^4$~K) and a ``hot'' ($T >
3\times 10^4$~K) gas phase.
In practice most of the ``hot'' gas in our simulations is much hotter
($T \sim 10^6$~K) than our demarcation threshold, due to cooling times
being short at intermediate temperatures and densities.
The cold gas is in a rather dense state ($n_H \ga
10^{-3}\,\mbox{cm}^{-3}$), while the hot gas is more dilute.

We measure the amount of gas that has been accreted onto the
central disk during each of the intervals $[t_i, t_i+\Delta t_i]$
between successive snapshots in our simulations.
For the runs presented in this work the snapshots are evenly spaced
with $\Delta t_i=1\,\mbox{Gyr}$.
The accreted material is classified as ``hot'' or ``cold'' according
to its state at~$t_i$.
A quantity of interest is the fraction~$\fhot$ of the total disk mass
that was in the form of hot gas just before being accreted onto the
existing cold disk.
In figure~\ref{f:jdisk-fhot} we plot the specific angular momenta 
$\jt_\disk \equiv { j_\disk / V_c^2 }$
of the disks in the final state of our simulations against~$\fhot$.
As one could have expected given the presence of an UVX background, 
$\fhot$ is rather large for all our simulations ($\fhot \ga 0.5$).
The simulations with reheating tend to have even higher values $\fhot
> 0.75$, reflecting the fact that the formation of cold clumps is
suppressed at early times.
While there is a clear trend in that on average the $BO$ runs have
both larger $\fhot$ and larger $\jt_\disk$ than the $RH$ runs, and the
$RH$ runs than the $PA$ runs, the scatter in $\jt_\disk$ vs.~$\fhot$
in each group of simulations is quite large. Within each group there
appears to be no correlation between $\jt_\disk$ and~$\fhot$.
This reminds us that factors other than the thermodynamic state
(density and temperature) of the accreted material play a role in
determining the angular momentum of the disk.

Let us examine the history of angular momentum acquisition.
In each time bin $[t_i, t_i+\Delta t_i]$ we compute the angular
momentum of both the infalling material ($\vec\jmath_{\infall,i} 
= j_{\infall,i}
\vec e_{\infall,i}$, where the $\vec e$'s are unit vectors) and the
preexisting disk (the unit vector of which at time~$t_i$ will be
written~$\vec e_{\disk,i}$).
In figure~\ref{f:jdisk-cos}
 we plot the average $\langle\vec e_{\disk,i} \cdot \vec
e_{\infall,i}\rangle$ over all time bins~$i$ for which data are
available.
The results are insensitive to whether only the hot gas (filled
symbols) or all the accreted gas (open symbols) is used in
determining~$\vec e_{\infall}$.
As expected, the final angular momentum is larger in disks that have
more coherent infall.
Interestingly, the $BO$ runs appear to possess this property of
experiencing coherent infall to a larger degree than the other sets of
runs. The reason, we believe, is that much
of the infalling material is material that has been ``blown out'' of a
clump at some earlier time and retains the imprint of the orbital
motion of its parent clump. In the $PA$ runs, the same material would
be accreted in a single event and cause an abrupt change in the
angular momentum of the disk. In the $BO$ runs, it is accreted over a
longer period and thus $\vec e_{\disk}$ already includes part of the
realignment. 

Figure~\ref{f:jdisk-sdcos} shows the distribution of specific angular
momenta~$\jt_\disk$ as a function of the dispersion $\sigma_{\vec
e_{\infall}}$ in the angular momentum orientation of the infalling
material. This dispersion is defined by
\begin{equation}
\sigma^2_{\vec e_{\infall}} = \langle \left(\vec e_{\infall,i} - \langle
\vec e_{\infall,i}\rangle\right)^2 \rangle,
\end{equation}
with both means being taken over the time bins~$i$, and measures the
scatter in the orientation of $\vec e_{\infall}$.
A perfectly isotropic distribution of $\vec e_{\infall}$ would
result in $\sigma_{\vec e_{\infall}}=1$.
This confirms the trend that the final angular momentum is
reduced when the infall is disordered.

In two of the $PA$ runs, the central object undergoes a major merger
of cold gas, defined as an event in which the mass $\Delta M_{\rm
cold}$ of cold gas
gained during the last Gyr is at least 30\% of the current disk mass~$M$.
These runs are the ones that possess the smallest specific angular momenta.
We plot the peak values of $\Delta M_{\rm cold}/M$ obtained during the
evolution of each of our simulated systems in
figure~\ref{f:jdisk-dmcold}.
The solid line on this plot shows the result of a linear least-squares
fit, and confirms the visual impression that there is a (negative)
correlation between major cold merger events and the final angular
momentum of the disk.

Figure~\ref{f:jdisk-jinf} shows $\jt_\disk$ versus the normalized total specific
angular momentum of the infalling gas, $\jt_\infall =
j_\infall/V_c^2$, defined by
\begin{equation}
j_\infall = { \left| \vec{J}_\disk(t_{i_0}) +  \sum_{i \ge i_0}
\vec{J}_\infall(t_i) \right| \over M_\disk(z=0) } .
\end{equation}
If no angular momentum were lost or gained by the disk other than by
accretion, we would expect $j_\disk$ to equal $j_\infall$.
This is not the case.
The difference indicates that a significant part of the angular
momentum is lost either during the last $\sim 0.5$~Gyr before
accretion or at the time of accretion.
This happens even when almost all the gas is accreted as hot and dilute:
the dependence of the angular momentum loss on the type
($PA$, $RH$, or $BO$) of model is much weaker than what we found in
figure~\ref{f:jdisk} for~$\jt_\disk$. 
We suspect that this late loss of angular
momentum is due to numerical transport effects as the SPH particles
get incorporated into the disk. If so, this contribution to the
angular momentum loss (a factor of about~4) is unlikely to have a
physical counterpart.
We should therefore consider our deficit of a factor of~5 in
$\jt_\disk$ for the $BO$ simulations (figure~\ref{f:jdisk}), and the
similar result of WEE98, as a success in that no further
unknown physical processes seem to be required.
If instead of $\jt_\disk$ (as in figure~\ref{f:jdisk}) we plot $\jt_\infall$
(figure~\ref{f:jinf}), the agreement in disk angular momentum between the
simulated and observed systems becomes quite satisfactory for all
of the $BO$ runs and even for some $RH$ and $PA$ runs.

One may wonder whether the large differences in $j_\infall$ between
runs started from the same initial conditions is due to their somehow
having accreted different material.
To the extent that there are differences in the final disk masses
$M_\disk$ (table~\ref{t:mrv}), this must clearly be the case; but as we have
seen, the variation in $M_\disk$ does not show any discernible
systematic trend.
Since most of the angular momentum is imparted by tidal torques in the
course of the simulations, it is not very instructive to compute the
initial angular momentum of the gas that ends up in the disk. Instead,
we evaluate the analogue of $j_\infall$ for the dark matter initially
associated with this gas, and ask whether this new quantity
\begin{equation}
j_{\infall,\dm} \equiv { \left| \vec{J}_{\disk,\dm} (t_{i_0}) +
\sum_{i \ge i_0} \vec{J}_{\infall,\dm}(t_i) \right| \over
M_{\disk,\dm}(z=0) }
\end{equation}
is in any way correlated with~$j_{\infall}$.
We plot $\jt_{\infall}$ against $\jt_{\infall,\dm}$ in figure~\ref{f:jinf-jinfdm}.
We find no evidence of a correlation.
The same can be said of $\jt_{\disk}$ versus~$\jt_{\infall,\dm}$ in
figure~\ref{f:jdisk-jinfdm}.

The connection between $j_\infall$ and~$j_{\infall,\dm}$ may of course
have been washed out by angular momentum exchange between the DM
particles in the halo, as well as by the divergence of orbits between
DM and gas particles that were initially associated with each other.
It is therefore equally instructive to compare $\jt_\infall$ to the
dimensionless spin parameter $\lambda$ of the final dark matter halo,
computed both within~$r_{200}$ and within the baryonic
infall radius~$r_{\infall}$. 
We do this in figure~\ref{f:jdisk-lambda}, and again find no signal.

Figure~\ref{f:jinfdm} shows $\jt_{\infall,\dm}$ versus circular speed $V_c$. As
proposed by FE80, tidal torques appear to produce just enough specific
angular momentum in the dark matter halos to be consistent with observations.
This, however, 
requires that the gas associated with the dark matter contracted 
dissipatively to form galactic disks with at most a minor loss of
angular momentum.
Figure~\ref{f:jinfdm-lambda} shows $\jt_{\infall,\dm}$
(computed for the DM associated with all infalling gas, both hot and cold)
versus the spin parameter 
$\lambda$ of the dark matter halo evaluated at the baryonic infall radius. 
A trend of increasing $\jt_{\infall,\dm}$ with increasing
$\lambda$ is seen, supporting the proposal by FE80 that the spin parameter
of a dark matter halo is an important parameter in determining the
specific angular 
momentum of the disk galaxy formed in its central region.

\section{Conclusions}
\label{s:conclusions}

We have carried out {\TSPH} simulations of the formation of Milky Way-sized
disk galaxies in a cosmological context.
For the ``passive'' variant of such simulations, i.e. when 
the effects of stellar feedback processes are neglected, we confirm
previous reports that the simulated disks are far too small (by about
a factor of~10) and poor in specific angular momentum (by a factor of
about~25) by comparison with observations of real galaxies.
This is true even for simulations where 80\% or more of the final
mass of the disk is accreted from hot, dilute gas rather than mergers
of cold clumps.
These ``passive'' simulations are a natural extension of the work of
VHSL94, and by using cosmological initial conditions rather than 
isolated uniform spheres they remove an objection raised by NS97
against that work.

Uniform reheating of all the gas in the Universe at $z\sim 6$, while
reducing the angular momentum deficit to some extent, is not as
effective as later blow-out of gas from a first generation of cold
star-forming clouds, at least as long as the reheating temperature
does not exceed $5\times 10^5$~K, which we estimate is a firm upper limit.
This estimate is based mainly on chemical enrichment constraints,
but it is worth noting that
with some (conservative) assumptions on the initial mass function,
one can get a similar bound from the mass of our Galaxy's stellar
halo.

Instantaneous blow-out (at $z\simeq 2.4$ in our simulations), on the
other hand, leads to a significant closing of the gap between observed
and simulated values of the specific angular momentum.
The remaining mismatch develops mostly in the late stages of gas
infall and in the subsequent evolution of the disk, suggesting that it
may be an artifact of artificial viscosity in the SPH method.

The success of the blow-out model appears related to the fact that the
gas is accreted onto the central disk in a very gentle way, without
abrupt changes of angular momentum orientation. There is no obvious
difference in the angular momentum of the dark matter associated
with the accreted gas between the different sets of simulations.

There is still considerable room for improvement in our ability to
simulate the formation of disk galaxies. Some of it will depend on our
ability to increase the numerical resolution. We have seen that the
structure of the disks is affected by gravitational softening, and
that much of the 
angular momentum is lost in the late phases of accretion. We plan to
discuss this issue in detail in a future paper; for now let us point
out that the test of shear-free artificial viscosity presented in NS97
is incomplete, in that it only considers the abrupt collapse of cold
gas, not the steady growth of the disk in a cooling flow (which would
be much more relevant to our simulations, and is also more in
keeping with the widely held view that actual galactic disks formed
by gradual accumulation of material). Irregularities in the spatial
arrangement of the SPH particles in the transition region from the hot
to the cold phase can also contribute to angular momentum transport, and
shear-free artificial viscosity formulations are more vulnerable to
such small-scale disorder (in effect, the flow is more turbulent).
While shear-free viscosity is undoubtedly helpful in preventing
further secular evolution within an already settled disk, increased
numerical resolution may be the only truly effective way of improving the
solution in the transition region from the hot halo to the cold disk.

It would also be very tempting to incorporate star formation into the
simulations in a more direct way, as other authors have done, by
converting some of the gas mass into new ``stellar'' collisionless
particles whose orbits are then integrated explicitly. To be
consistent with observations of the stellar content of our own Galaxy,
the bulk of this conversion should only occur after the gas has
settled into the disk.
A natural expectation is that this conversion might allow more of the
angular momentum in the disk to be retained, since the stellar
component is not subject to artificial viscous effects.
However, the results of Steinmetz \& Navarro (\cite{SN98}) suggest
to us that this effect is not sufficiently strong to make a
significant difference to the results, at least in the situation they
considered (which did not include any particular mechanisms for
delaying gas cooling and star formation).

We have benefited considerably from the comments of
K. Gorski, U. Hellsten, C. Lacey, J. Navarro, B. Pagel, an anonymous
referee, and the editor S. Shore.
This work was supported by Danmarks Grundforskningsfond through its support
for the establishment of the Theoretical Astrophysics Center.

\newpage
\figcaption[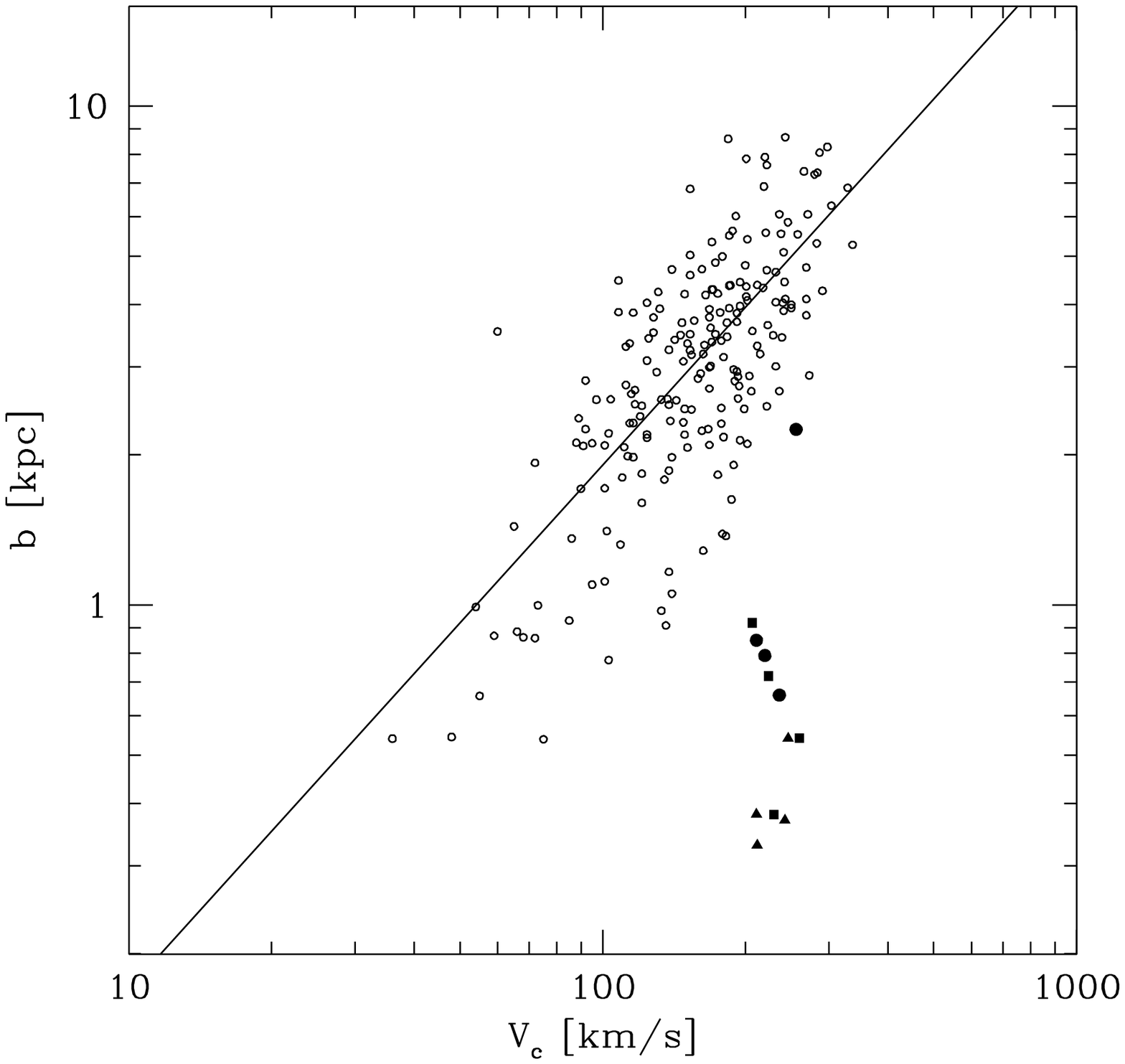]{Exponential disk scale lengths $b$ as a function of circular
speed $V_c$ for a sample of 203 Sb--Sd galaxies (open circles), and for the
disks in our $PA$ (``passive'', triangles), $RH$ (``reheating'',
squares), 
and $BO$ (``blow-out'', circles)
simulations. The solid line is a linear least-squares fit to the
observational data.\label{f:b-vc}}

\figcaption[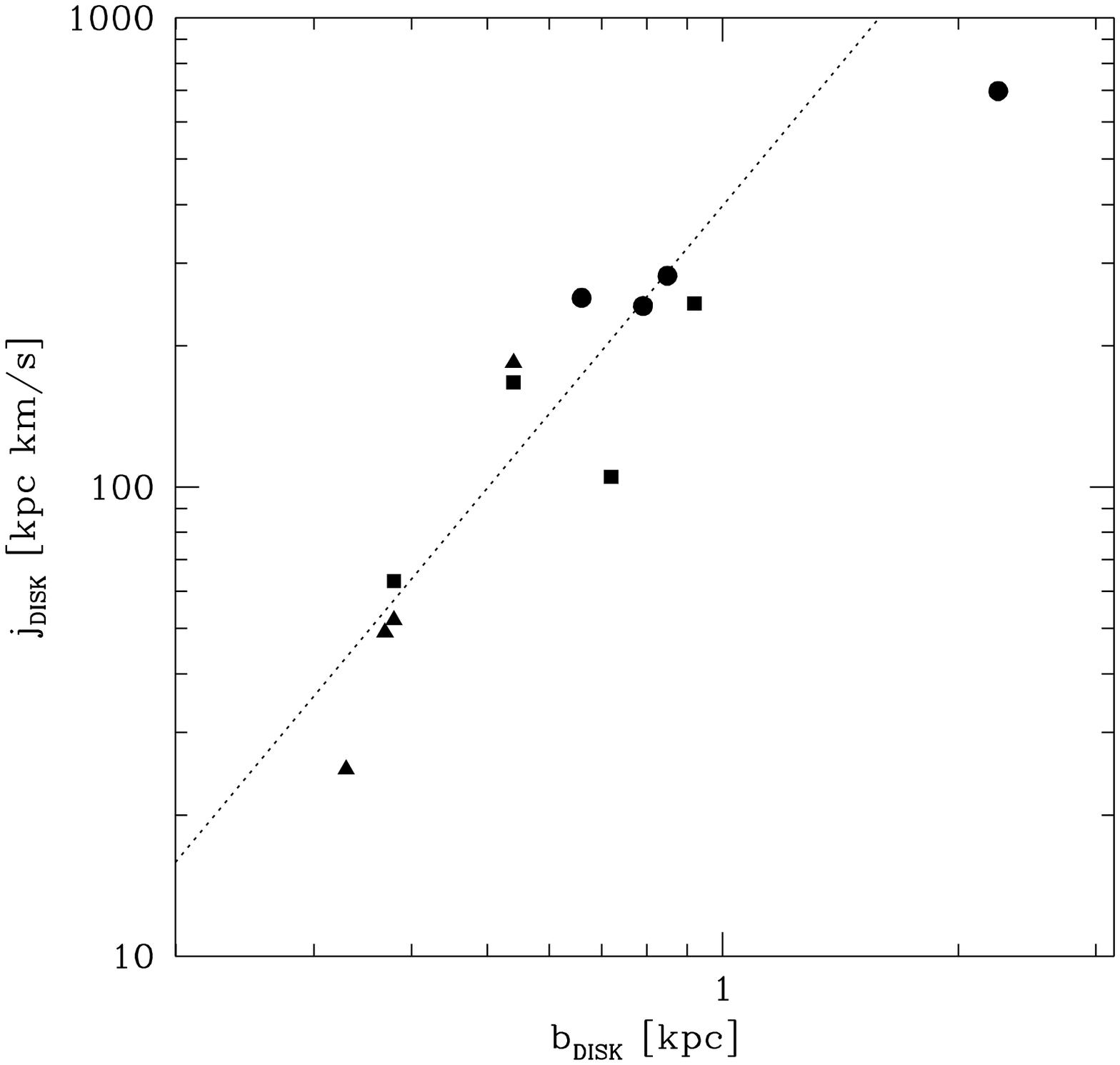]{Specific angular momenta versus scale length for the
simulated disks (symbols as in figure~\ref{f:b-vc}).
The dotted line has a slope of~2, corresponding to $j_\disk \propto
b_\disk^2$ (see text), and an arbitrary zero point. \label{f:j-b}}

\figcaption[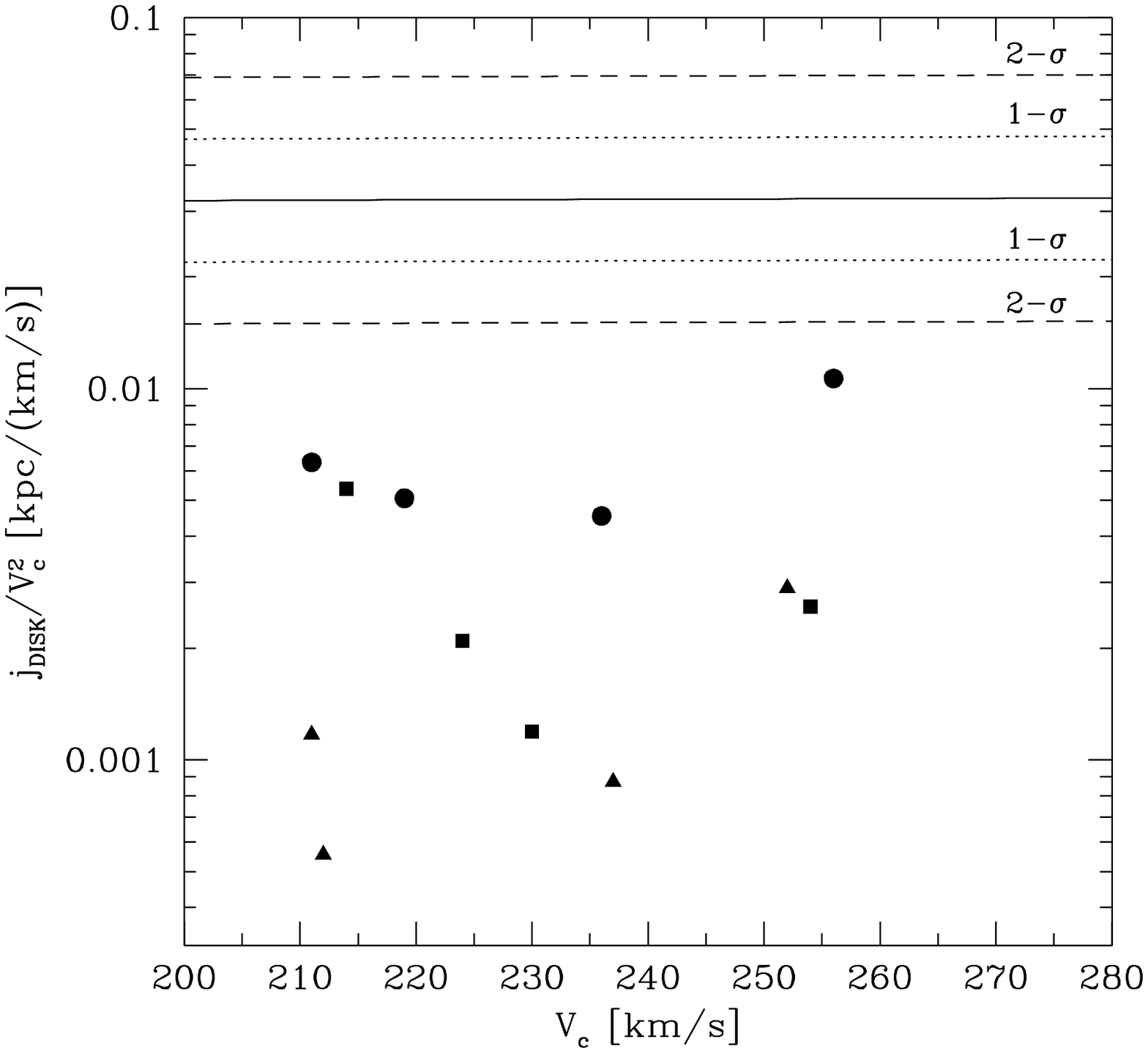]{Normalized specific angular momentum $\jt_\disk \equiv
j_\disk/V_c^2$ for the simulations (symbols as in
figure~\ref{f:b-vc}). The solid line shows the mean value from the
observational data, the dotted and dashed lines bracket the $1\sigma$
and $2\sigma$ intervals around this mean. \label{f:jdisk}}

\figcaption[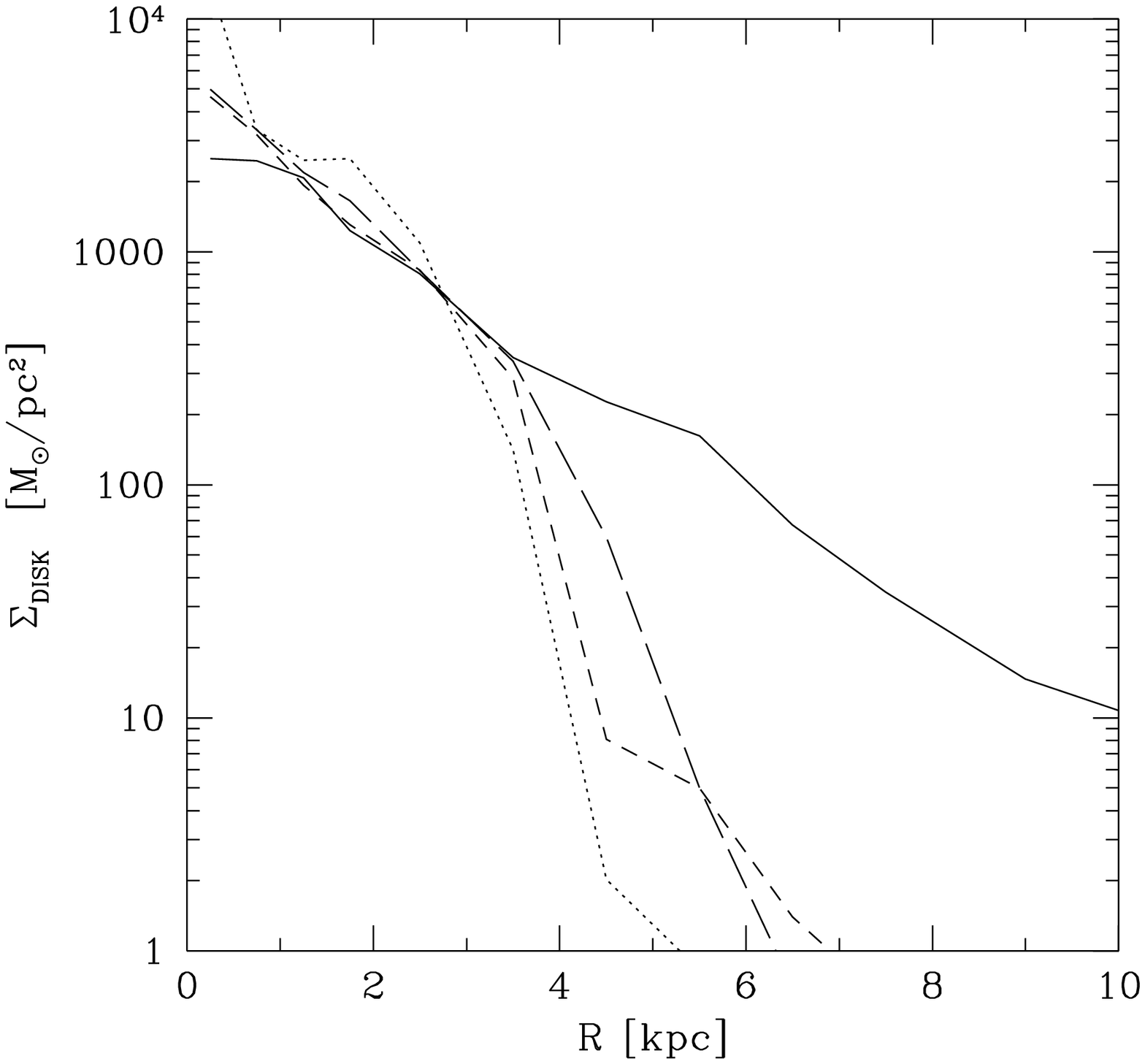]{Azimuthally averaged disk surface density profiles at
$z=0$ of model galaxies
$S1_{BO}$ (long-dashed line), $S2_{BO}$ (short-dashed line), $S3_{BO}$ (dotted
line) and $S4_{BO}$ (solid line) from the gas ``blow-out''
simulations.
\label{f:sd-r-bo}}

\figcaption[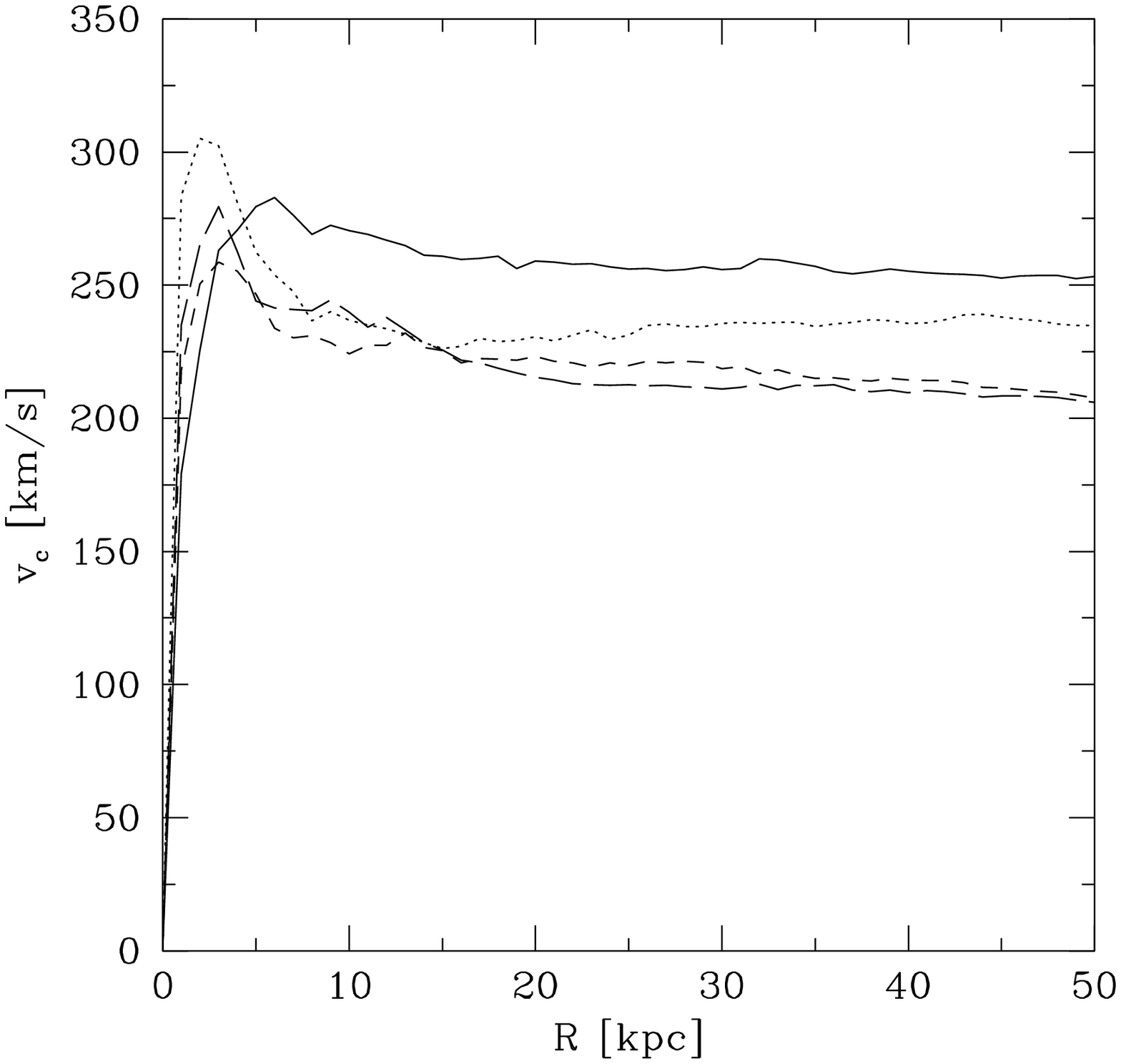]{Rotation curves at $z=0$ for the ``blow-out'' models.
Symbols are as in figure~\ref{f:sd-r-bo}. \label{f:vc-r-bo}}

\figcaption[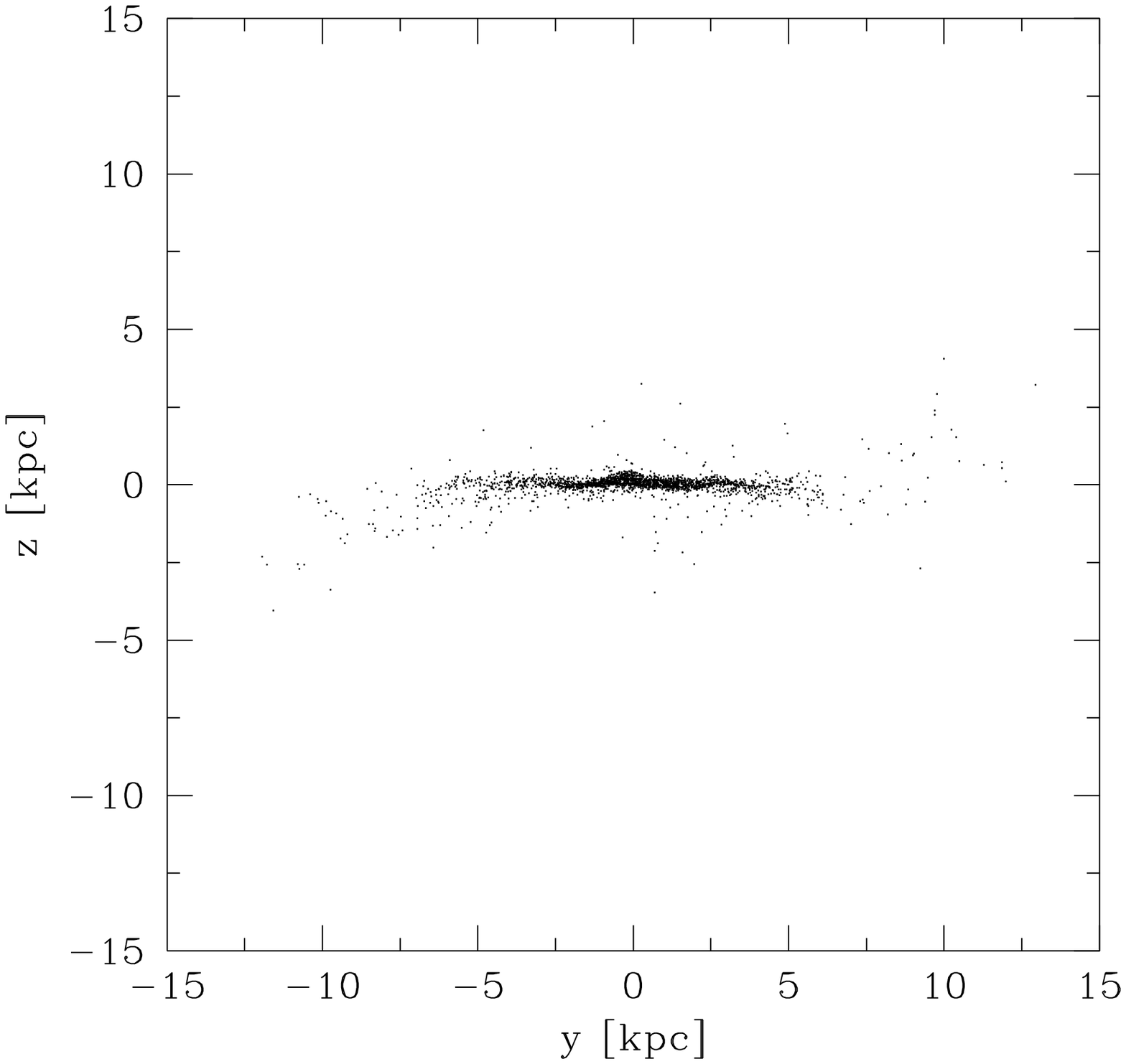]{Edge-on view of model galaxy $S4_{BO}$ at
$z=0$. \label{f:z-y-s4bo}}

\figcaption[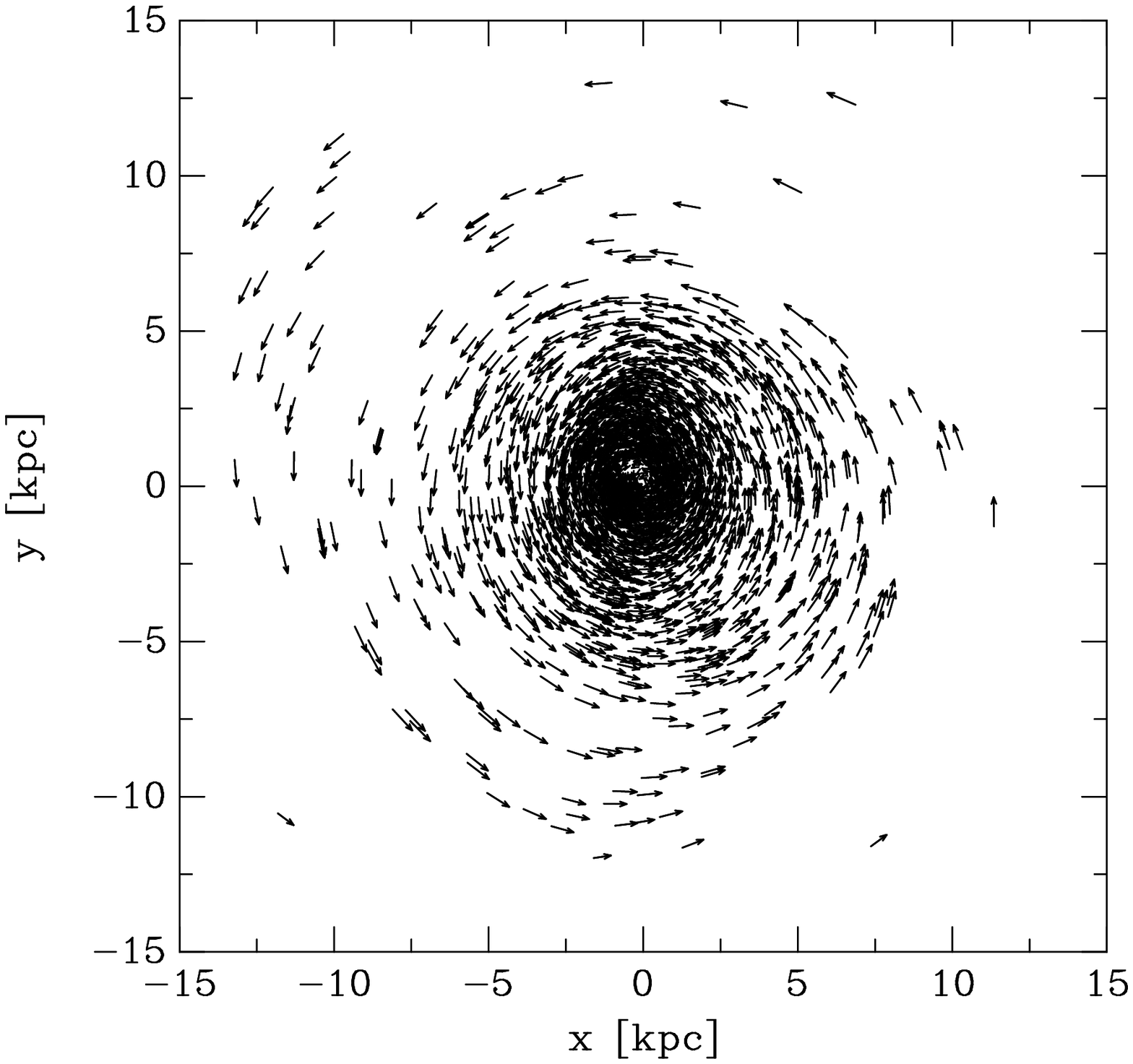]{Face-on view of model galaxy $S4_{BO}$ at
$z=0$. Arrows represent the velocities of individual gas particles. 
\label{f:y-x-s4bo}}

\figcaption[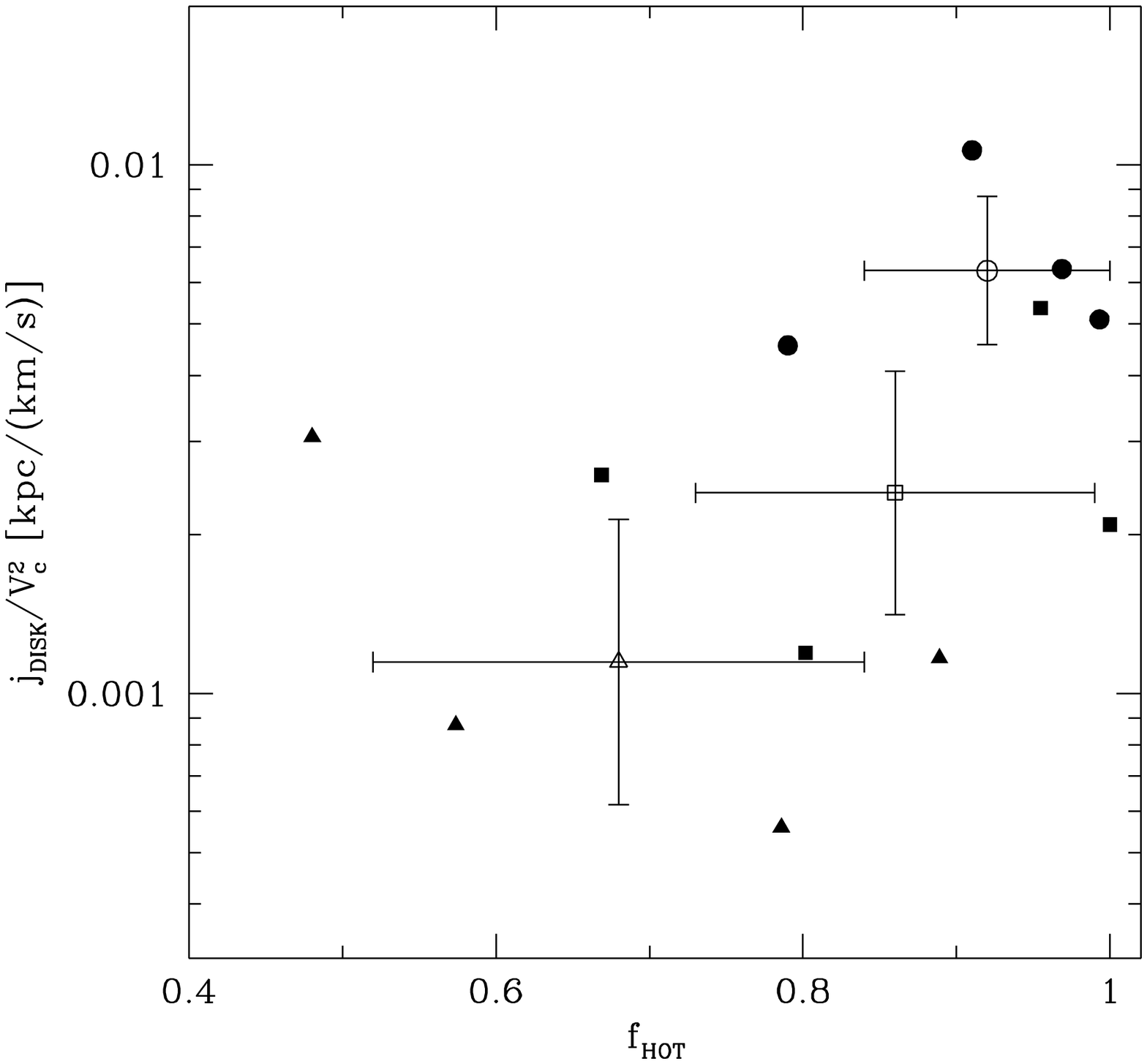]{$\jt_\disk \equiv j_\disk/V_c^2$ versus the hot gas
infall fraction $\fhot$ for the simulations (symbols as in
figure~\ref{f:b-vc}).
The open symbols with error crosses represent the mean and standard
deviation within each set of four simulations. There is no clear
correlation between $\jt_\disk$ and $\fhot$ within each set.
\label{f:jdisk-fhot}}

\figcaption[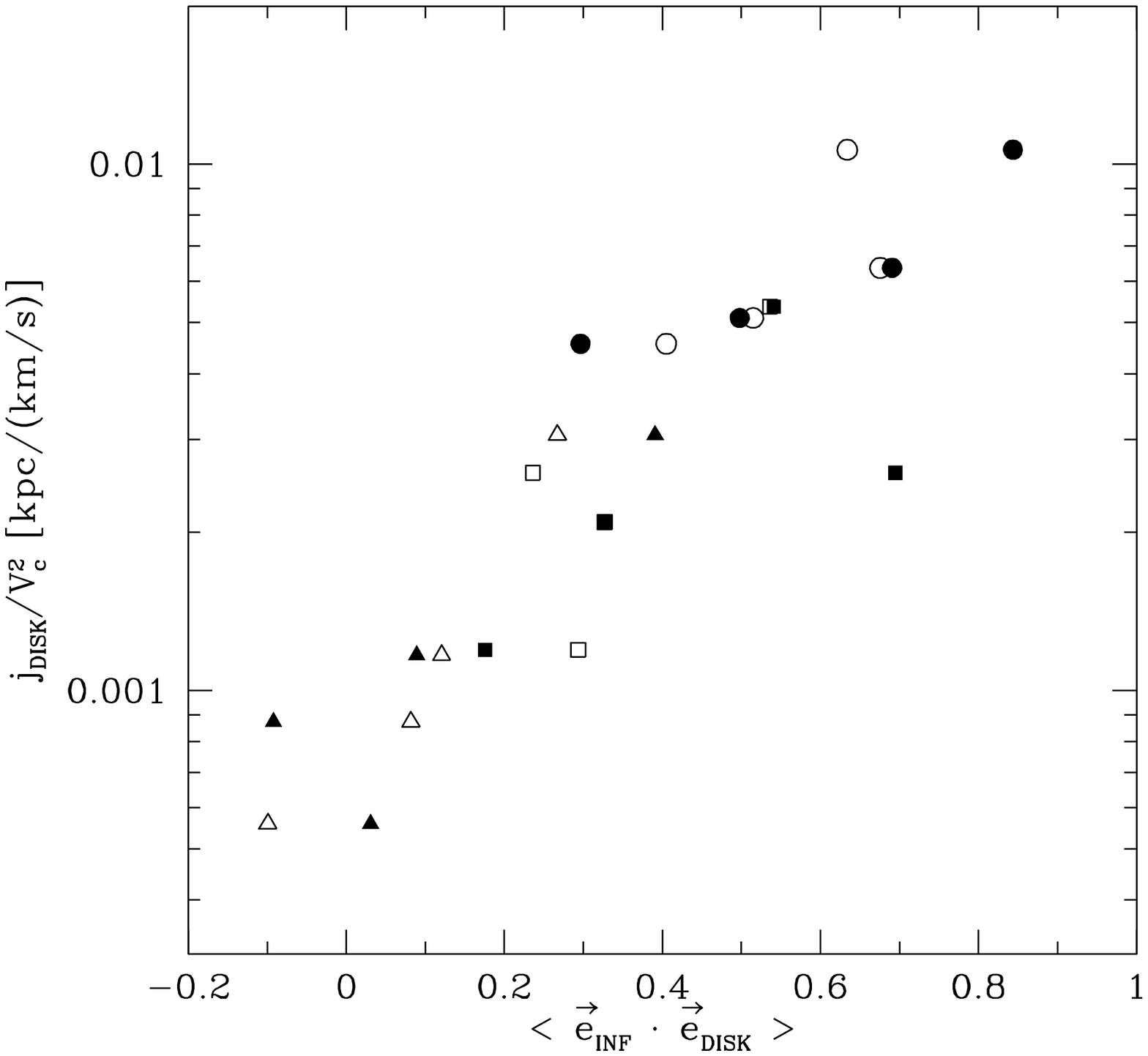]{$\jt_\disk$ versus the mean cosine $\langle
\vec{e}_\infall \cdot \vec{e}_\disk \rangle$ for the
simulations. Filled symbols show results obtained by including only
the hot gas in the definition of $\vec{J}_\infall$. Open symbols show
results obtained with all gas (hot and cold). Symbol shapes as in
figure~\ref{f:b-vc}. \label{f:jdisk-cos}}

\figcaption[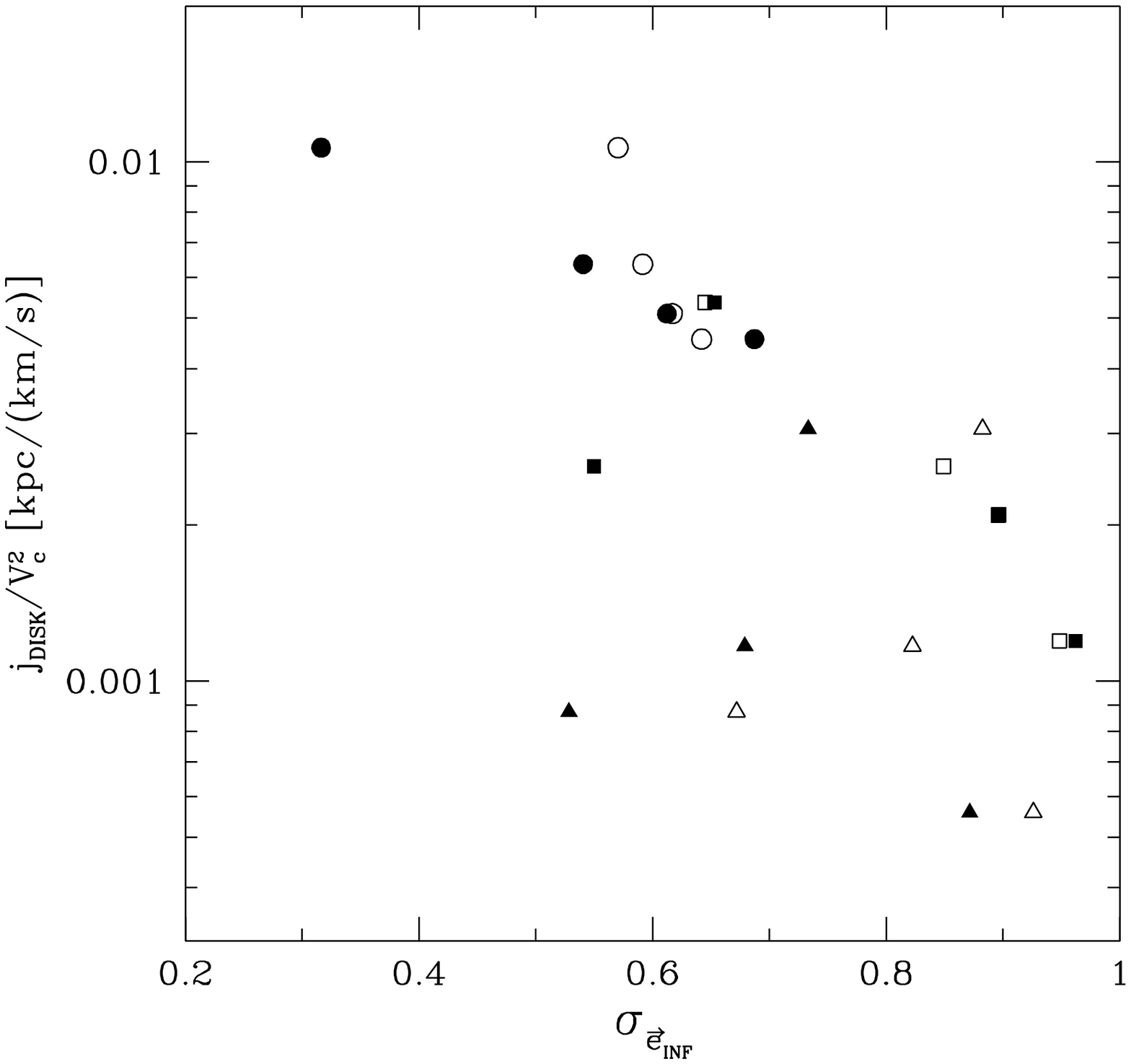]{$\jt_\disk$ versus the dispersion in the cosines of
figure~\ref{f:jdisk-cos}. Symbols as in figure~\ref{f:jdisk-cos}.
\label{f:jdisk-sdcos}}

\figcaption[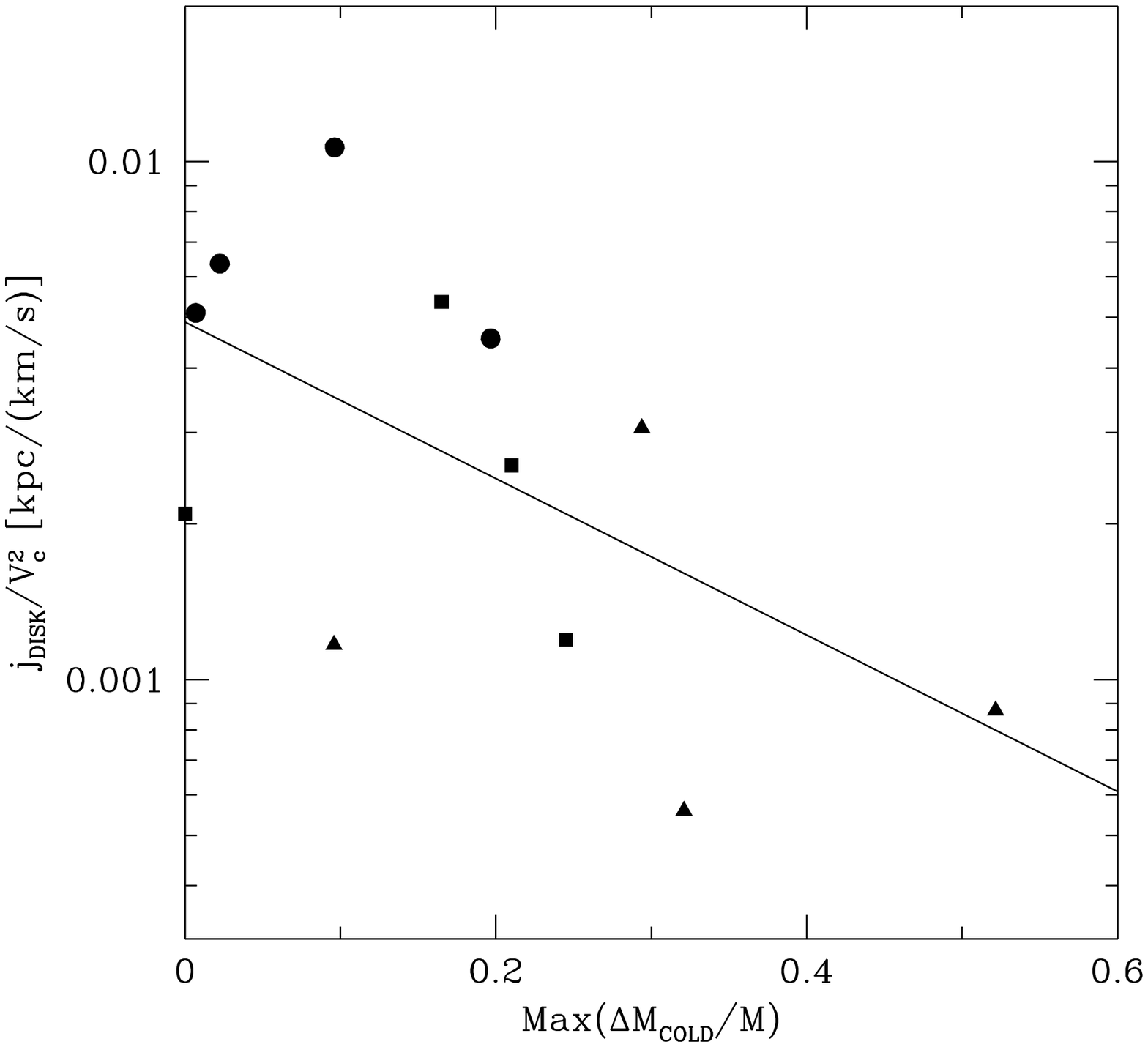]{$\jt_\disk$ versus the maximum cold merger strength
$\max\left(\Delta M_\cold/M\right)$ for the simulations. Symbols are
as in figure~\ref{f:b-vc}. The solid line is a linear least squares
fit.
\label{f:jdisk-dmcold}}

\figcaption[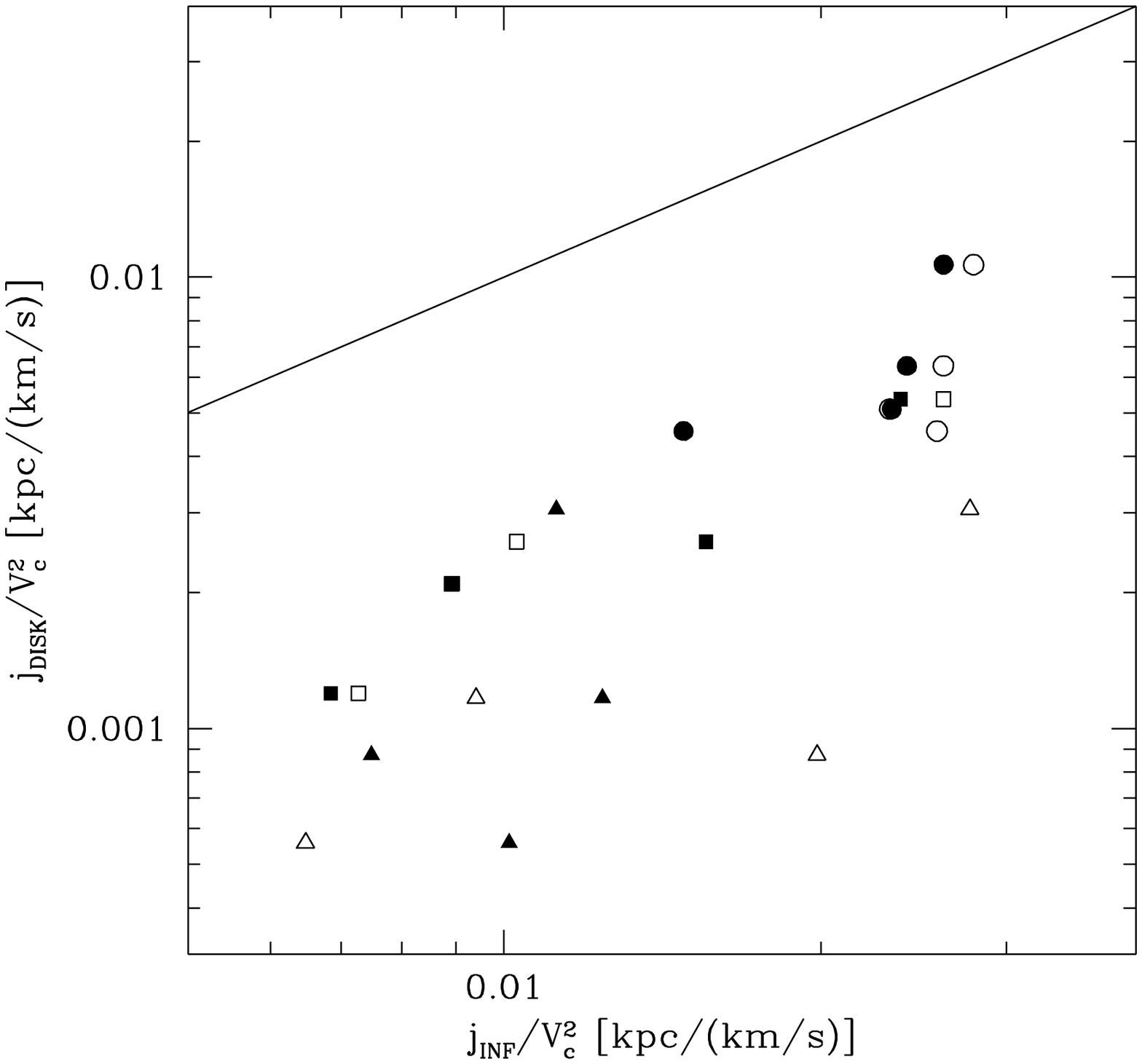]{$\jt_\disk$ versus $\jt_\infall$ for the simulations.
Symbols as in figure~\ref{f:jdisk-cos}. \label{f:jdisk-jinf}
The solid line corresponds to $\jt_\disk = \jt_\infall$.}

\figcaption[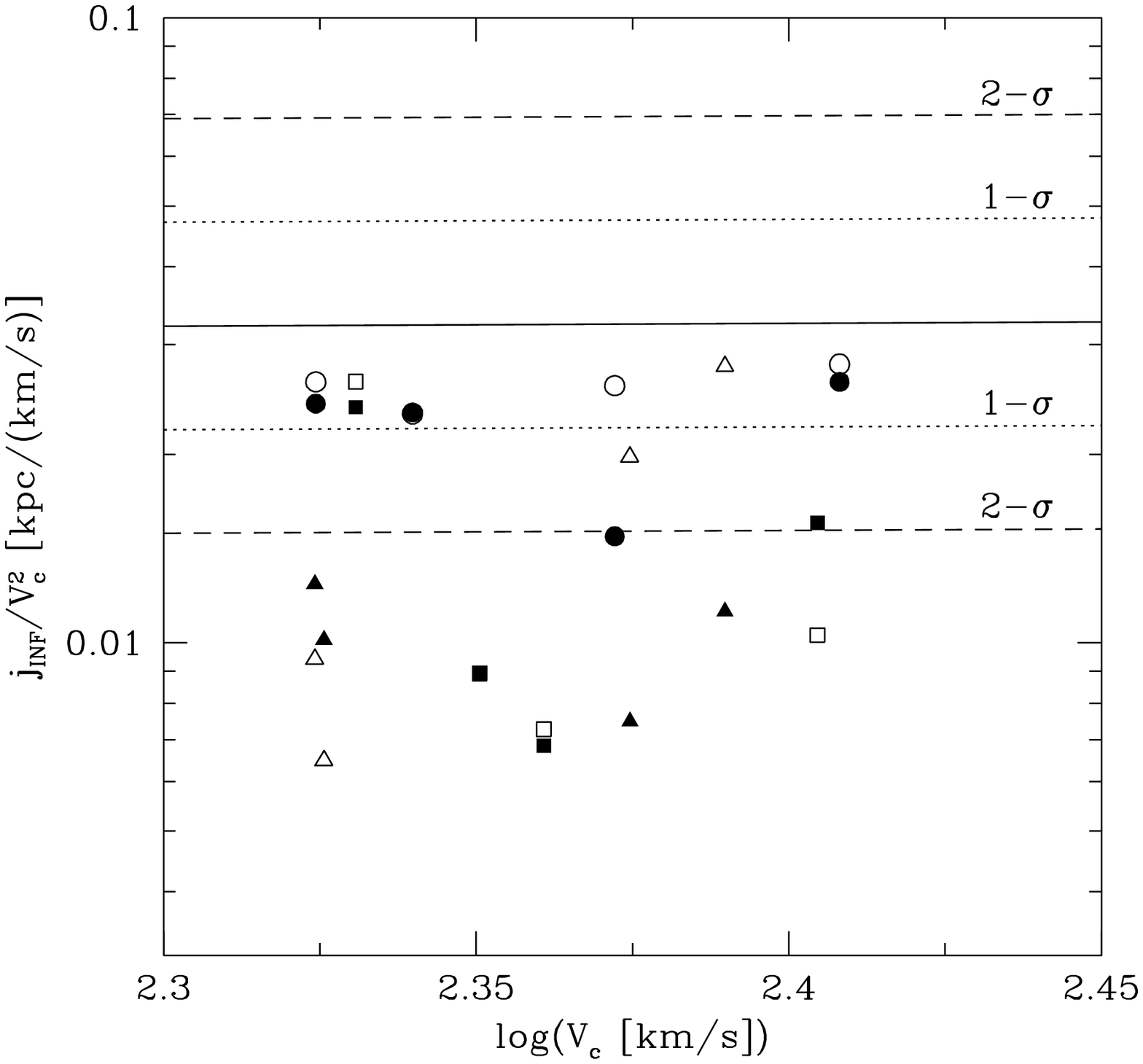]{$\jt_\infall$ for the simulations. Symbols as in
figure~\ref{f:jdisk-cos}, horizontal lines as in figure~\ref{f:jdisk}.
\label{f:jinf}}

\figcaption[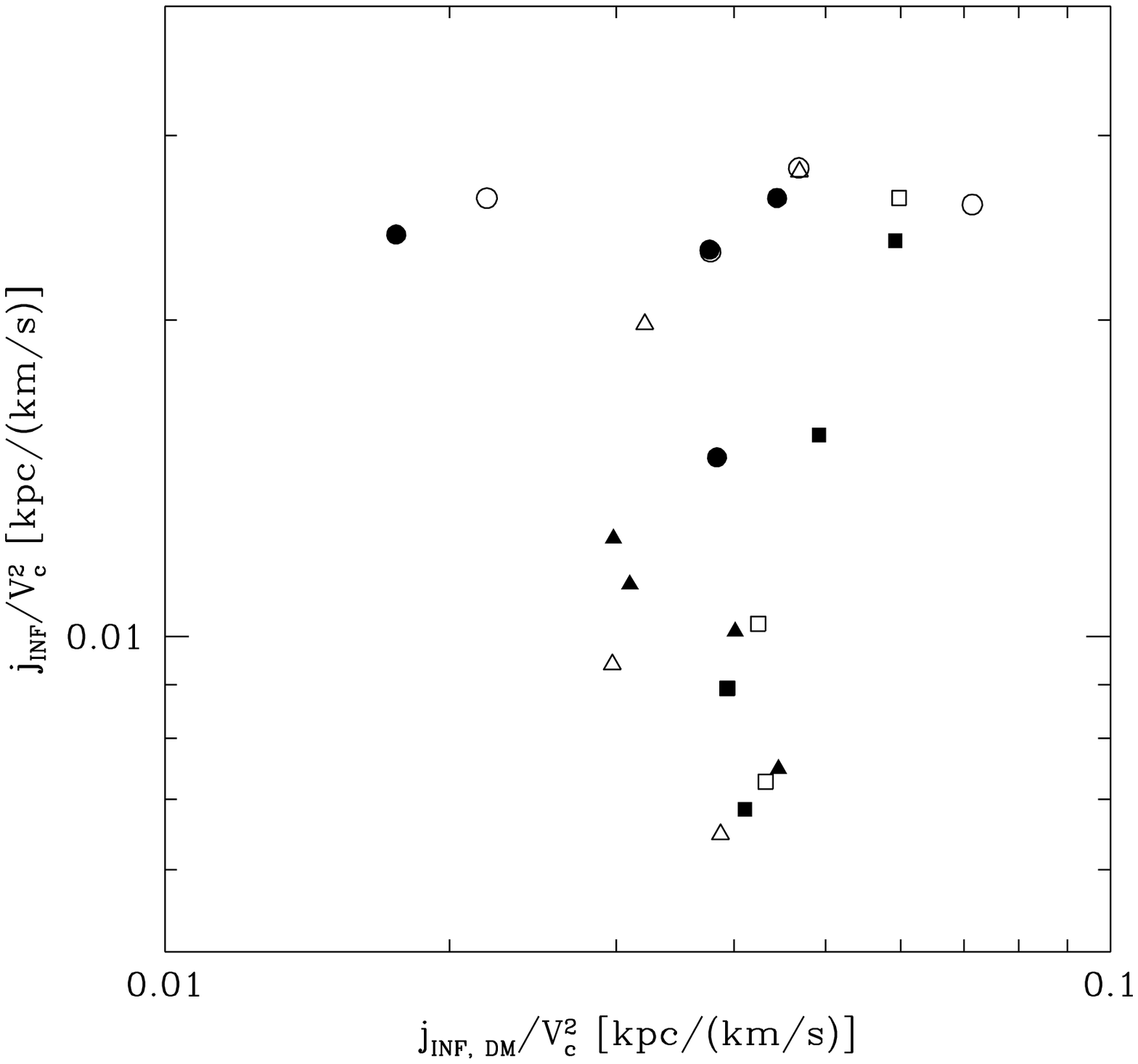]{$\jt_\infall$ versus $\jt_{\infall,\dm}$ for the
simulations.
Symbols are as in figure~\ref{f:jdisk-cos}.
\label{f:jinf-jinfdm}}

\figcaption[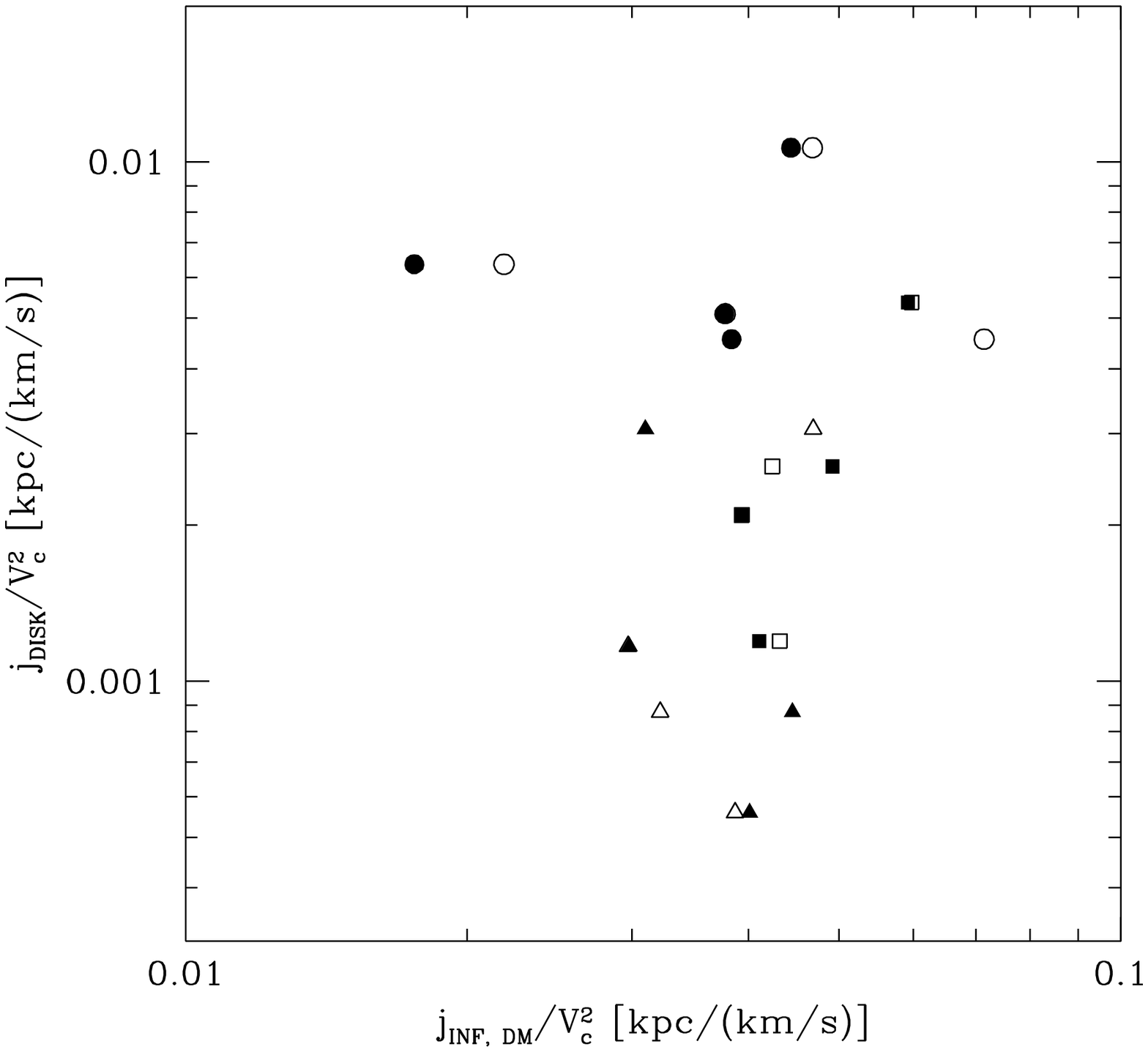]{$\jt_\disk$ versus $\jt_{\infall,\dm}$ for the
simulations.
Symbols are as in figure~\ref{f:jdisk-cos}.
\label{f:jdisk-jinfdm}}

\figcaption[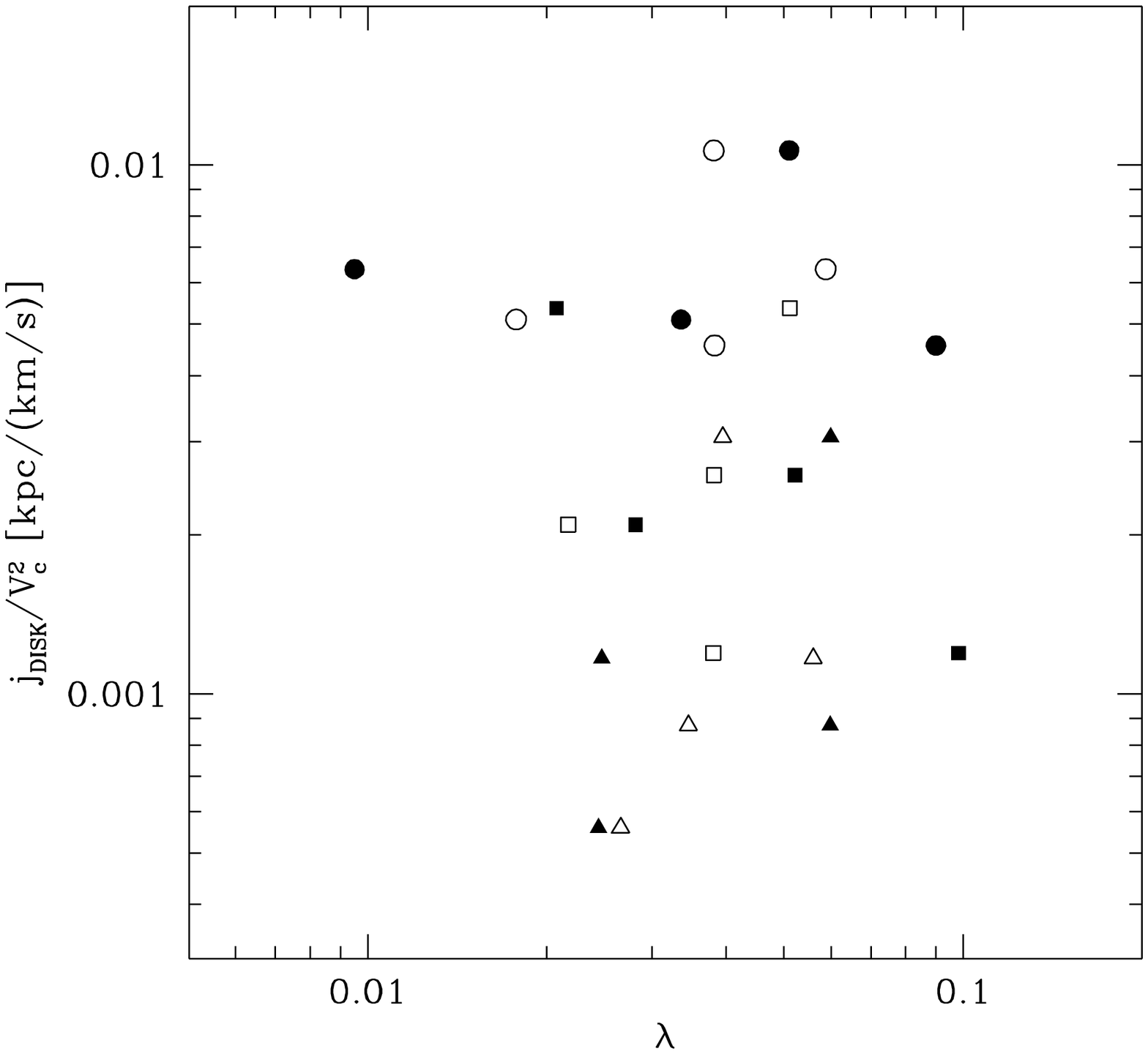]{$\jt_\disk$ versus the spin parameter $\lambda$. Filled
symbols are for $\lambda$ evaluated at the baryonic infall radius
$r_\infall$, open symbols for $\lambda$ evaluated at the virial radius
$r_{200}$. Symbol shapes are as in figure~\ref{f:b-vc}.
\label{f:jdisk-lambda}}

\figcaption[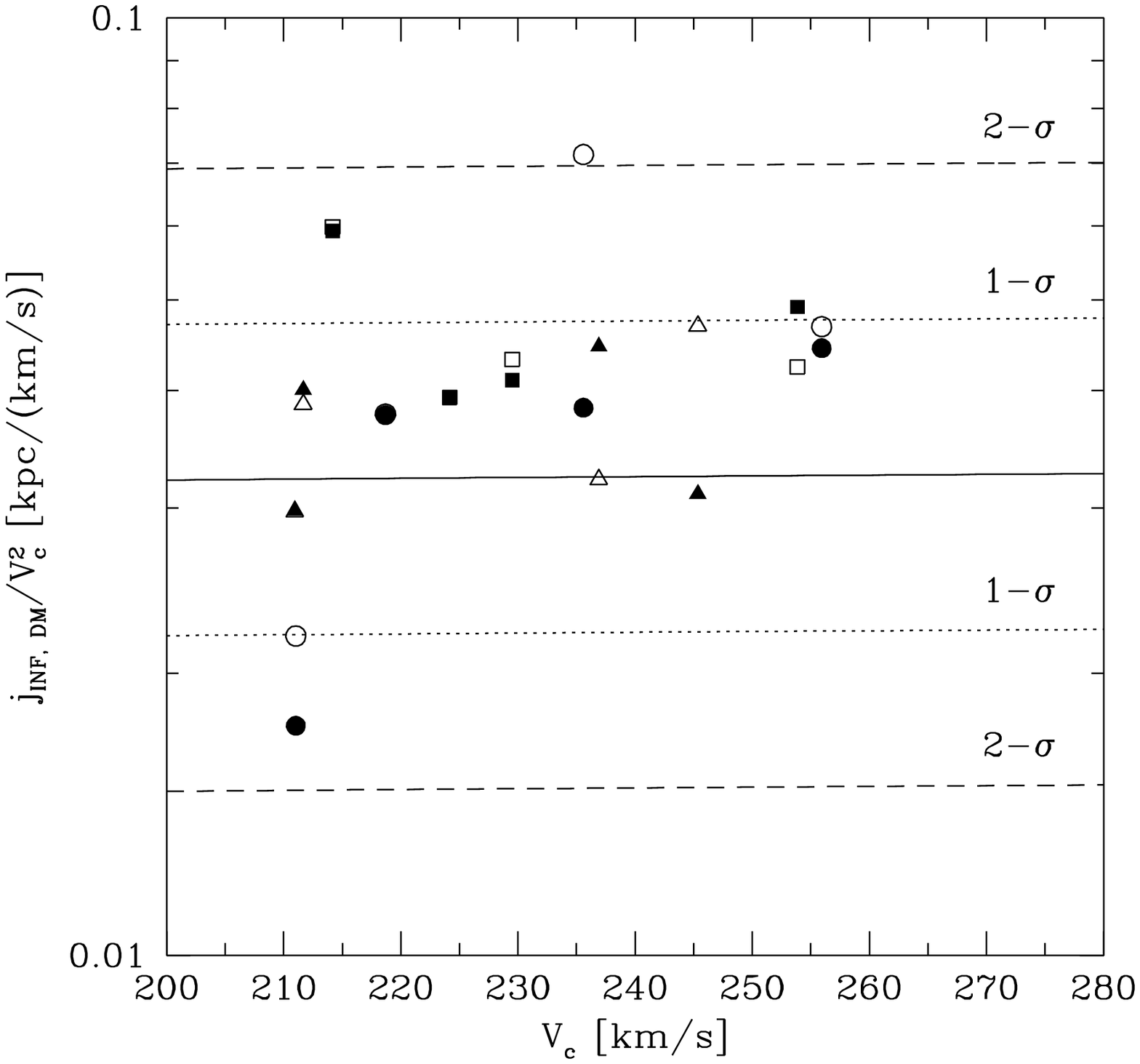]{$\jt_{\infall,\dm}$ for the simulations.
Symbols as in figure~\ref{f:jdisk-cos}, lines as in
figure~\ref{f:jdisk}.
\label{f:jinfdm}}

\figcaption[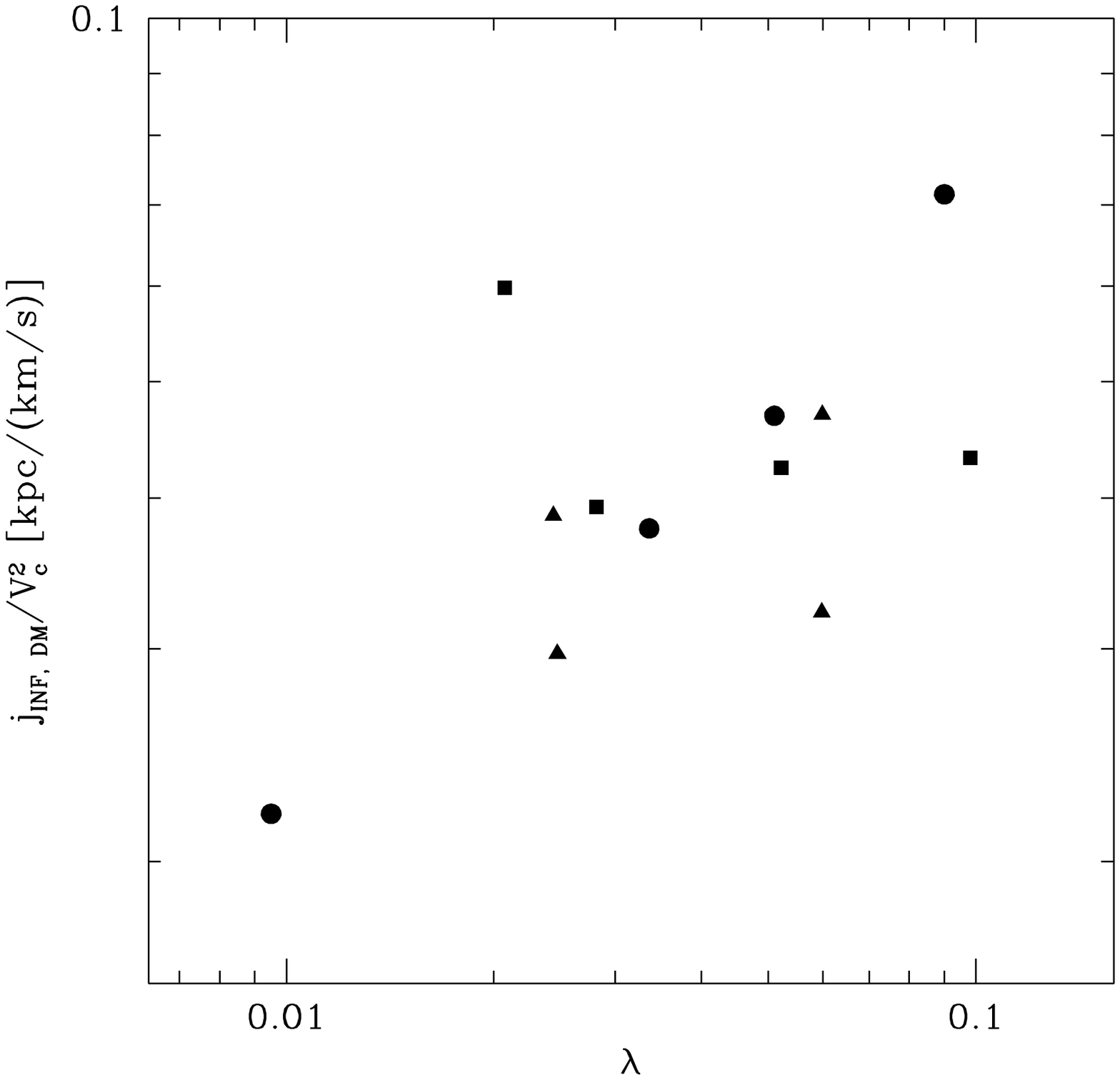]{$\jt_{\infall,\dm}$ versus spin parameter
$\lambda(r_\infall)$. Symbols as in figure~\ref{f:b-vc}.
\label{f:jinfdm-lambda}}

\clearpage
\newpage        
\begin{table}
\caption{Particle counts in the high-resolution spheres \label{t:np}}
\begin{tabular}{lrr}
\hline\hline
Label  & $N_{\rm SPH}$ & $N_{\rm DM}$\\
\hline
S1 & 13712 & 6856\\
S2 & 14170 & 7085\\
S3 & 14436 & 7218\\
S4 & 14726 & 7363\\
\hline \hline
\end{tabular}
\end{table}

\begin{table}
\caption{Masses, sizes and velocities at $z=0$ \label{t:mrv}}
\small
\begin{tabular}{lcccccccccc}
\hline\hline
Label  & $M_{200}$ & $r_{200}$ & $V_{200}$ & $N_{gas}$ & $N_{DM}$ &
$M_{gas}$ & $M_{DM}$ & $N_{disk}$ & $M_{disk}$ & $M_{disk}$ \\
                         & [10$^{12}$ M$_{\odot}$]  & [kpc] & [km~s$^{-1}$] & & &
                              [10$^{10}$ M$_{\odot}$]  & [10$^{12}$ M$_{\odot}$] & &    
                              [10$^{10}$ M$_{\odot}$]  & [$\Omega_b M_{200}$]       \\
\hline
S1$_{PA}$ & 1.88 & 318 & 159 & 2937 & 1651 & ~~8.40 & 1.80 & 1574 & ~~4.50 & 0.48\\
S2$_{PA}$ & 2.51 & 350 & 176 & 3922 & 2212 & 11.22 & 2.40 & 1566 & ~~4.48 & 0.36\\
S3$_{PA}$ & 2.53 & 331 & 181 & 4285 & 2218 & 12.26 & 2.41 & 2273 & ~~6.50 & 0.51\\
S4$_{PA}$ & 3.35 & 385 & 193 & 5418 & 2930 & 15.50 & 3.19 & 3260 & ~~9.33 & 0.56\\
S1$_{RH}$ & 2.08 & 327 & 165 & 3416 & 1819 & ~~9.77 & 1.98 & 1521 & ~~4.35 & 0.42\\
S2$_{RH}$ & 2.66 & 355 & 180 & 4276 & 2337 & 12.23 & 2.54 & 1665 & ~~4.76 & 0.36\\
S3$_{RH}$ & 2.55 & 332 & 182 & 4232 & 2236 & 12.11 & 2.43 & 1774 & ~~5.08 & 0.40\\
S4$_{RH}$ & 3.31 & 382 & 193 & 5310 & 2905 & 15.19 & 3.16 & 2896 & ~~8.29 & 0.50\\
S1$_{BO}$ & 2.10 & 328 & 166 & 3303 & 1847 & ~~9.45 & 2.01 & 1818 & ~~5.20 & 0.49\\
S2$_{BO}$ & 2.62 & 353 & 178 & 4038 & 2305 & 11.55 & 2.51 & 1606 & ~~4.59 & 0.35\\
S3$_{BO}$ & 2.58 & 352 & 178 & 4202 & 2267 & 12.02 & 2.46 & 2190 & ~~6.27 & 0.48\\
S4$_{BO}$
      & 3.34 & 385 & 193 & 5017 & 2946 & 14.35 & 3.20 & 2219 & ~~6.35 & 0.38\\ 
\hline \hline
\end{tabular}
\end{table}

\begin{table}
\caption{Angular momenta at $z=0$ \label{t:j}}
\begin{tabular}{lccccc}
\hline\hline
Label  & $V_c$ & $b_{disk}$ & $\lambda$ & $j_{disk}$ & $r_{inf}$\\
                         & [km~s$^{-1}$] & [kpc] & & [kpc km~s$^{-1}$] & [kpc]\\
\hline
S1$_{PA}$ & 212 & 0.33 & 0.024 & ~~25 & 112\\
S2$_{PA}$ & 211 & 0.38 & 0.025 & ~~52 & ~~83\\
S3$_{PA}$ & 237 & 0.37 & 0.060 & ~~49 & 137\\
S4$_{PA}$ & 252 & 0.54 & 0.060 & 184 & 138\\
S1$_{RH}$ & 214 & 0.92 & 0.021 & 246 & ~~99\\
S2$_{RH}$ & 224 & 0.72 & 0.028 & 105 & ~~98\\
S3$_{RH}$ & 230 & 0.38 & 0.098 & ~~63 & 100\\
S4$_{RH}$ & 254 & 0.54 & 0.049 & 167 & 123\\
S1$_{BO}$ & 211 & 0.84 & 0.010 & 282 & 121\\
S2$_{BO}$ & 219 & 0.80 & 0.034 & 243 & ~~94\\
S3$_{BO}$ & 236 & 0.66 & 0.090 & 253 & 120\\
S4$_{BO}$ & 256 & 2.25 & 0.051 & 698 & ~~87\\
\hline \hline
\end{tabular}
\end{table}

\end{document}